\documentclass[a4paper, 12pt]{article}
\usepackage{amssymb}
\usepackage{graphicx}
\usepackage[round]{natbib}
\usepackage{amscd}
\usepackage[ansinew]{inputenc}
\usepackage[english]{babel}
\usepackage{amsfonts}
\usepackage{dsfont}
\usepackage{latexsym}
\usepackage{array}
\usepackage{epsfig}
\usepackage{amssymb}
\usepackage[figuresright]{rotating}
\usepackage{soul}
\usepackage[centertags]{amsmath}
\usepackage{multirow}
\usepackage{lscape}
\usepackage{setspace}
\usepackage{ctable}
\usepackage{rotating}
\usepackage[a4paper,left=2.5cm,right=2.5cm,top=2cm,bottom=2cm]{geometry}
\usepackage{hyperref}
\usepackage{color}

\definecolor{eublue}{rgb}{0.1,0.1,0.5}
\newcommand{\bbeta}{\mbox{\boldmath$\beta$}}
\newcommand{\bphi}{\mbox{\boldmath$\phi$}}
\newcommand{\bgamma}{\mbox{\boldmath$\gamma$}}
\newcommand{\bdelta}{\mbox{\boldmath$\delta$}}
\newcommand{\bzeta}{\mbox{\boldmath$\zeta$}}
\newcommand{\bxi}{\mbox{\boldmath$\xi$}}
\newcommand{\bpi}{\mbox{\boldmath$\pi$}}
\newcommand{\bmu}{\mbox{\boldmath$\mu$}}

\begin{document}

\title{Omitted covariates bias and finite mixtures of regression models for longitudinal responses }

\author{Marco Alf\`o, Roberto Rocci\\ Dipartimento di Scienze Statistiche, Sapienza Universit\`a di Roma}

\maketitle
\onehalfspacing

\begin{abstract}

\noindent Individual-specific, time-constant, random effects are often used to model dependence and/or to account for omitted covariates in regression models for longitudinal responses. Longitudinal studies have known a huge and widespread use in the last few years as they allow to distinguish between so-called \emph{age} and \emph{cohort} effects; these relate to differences that can be observed at the beginning of the study and stay persistent through time, and changes in the response that are due to the temporal dynamics in the observed covariates. While there is a clear and general agreement on this purpose, the random effect approach has been frequently criticized for not being robust to the presence of correlation between the observed (i.e. covariates) and the unobserved (i.e. random effects) heterogeneity. Starting from the so-called \emph{correlated effect approach}, we argue that the random effect approach may be parametrized to account for potential correlation between observables and unobservables. Specifically, when the random effect distribution is estimated non-parametrically using a discrete distribution on finite number of locations, a further, more general, solution is developed. This is illustrated via a large scale simulation study and the analysis of a benchmark dataset.
\end{abstract}


\section{Introduction}
Individual-specific, time-constant, random effects are often used to model responses from a longitudinal study covering a limited, usually short, time window. As units participating in this kind of studies are often not randomly sampled from the corresponding reference population, and/or they are not randomly given treatment of interest, currently observed covariates may depend on the whole personal history and individuals may not be considered as homogeneous. Therefore, when modelling longitudinal data, we often face two key features: between individuals (unobserved) heterogeneity, and within-individual dependence. To help account for unobserved heterogeneity, individual-specific random effects are included in the linear predictor; since repeated measurements from the same individual share such common latent features, a (rather simple) structure of within individual dependence is also accounted for. This approach is especially suited for those empirical situations where a proper multivariate distribution is not available, e.g. for non Gaussian responses. The use of the so-called \emph{fixed effect} estimator is frequently advocated due to the claim that the approach based on random effects would not lead to a consistent estimator in the presence of dependence between the observed covariates and the random effects. As a by-product of the discussion, we may also notice that conditions for the fixed effect estimator to yield consistent estimates of individual-specific effects are seldom fulfilled. It is worth noticing that the two estimators may be based on the same working hypotheses and what really differentiates them is the use of a conditional rather than a marginal maximum likelihood approach, see \cite{mund:1978}. In this sense, the debate about the  \emph{fixed} vs \emph{random} nature of the individual-specific effects is misleading and makes no sense, see e.g. \cite{wool:2009} for a thorough discussion of the topic. Rather, we have to talk about individual-specific effects and approaches to estimation (and parameter interpretation as well) that can be made conditional or unconditional (marginal) to the individual effects. The Hausman test, see \cite{haus:1978}, is often adopted to choose which of the two approaches seem to be supported by the observed data. However, as it can be easily shown, it reduces to testing for a particular specification to hold for the random effect distribution; in fact, this is rather a working assumption that is not strictly necessary for the random effect approach to be used. 

To define a random effect estimator in the presence of correlation (generally dependence) between random effects  (unobserved heterogeneity) and observed covariates, we may consider the \emph{auxiliary equation} proposed by \cite{mund:1978} and later extended by \cite{cham:1980, cham:1982}, who defined the so-called \emph{correlated} random effect estimator. The topic has been discussed in the statistical world by several authors; just to cite a few, see e.g. \cite{neuh:kalb:1998}, \cite{neuh:mccu:2006}, \cite{rabe:skro:2014}, and quite recently and in  a very different context, \cite{lamo:etal:2016}. An interesting contribution to the debate has been given by \cite{kris:2005}, who discussed an approach based on the so-called \emph{QP decomposition}, discussed in the generalized linear models framework by \cite{neuh:kalb:1998}. It is worth mentioning that, however, the work by \cite{kris:2005} is seldom referenced, even if it shades some interesting lights on the \cite{mund:1978} approach. While all these alternatives provide reliable estimates, we think that a less parametrized approach, based on finite mixtures and a proper representation of the prior masses could be an efficient and general solution to the problem.

The paper is structured as follows. In section \ref{longdata}, we introduce longitudinal studies, adopted notation and basic issues of random effect models. In section \ref{aux} the QP decomposition is introduced in order to define a suitable parametrization of the random effects that allows consistent estimation of the within/between effects in the presence of correlation between the random effects and the observed covariates. In section \ref{mixing} we introduce an estimator of the random effect distribution based on the theory of nonparametric maximum likelihood estimation (NPMLe) of the mixing distribution. After discussing ML estimation in the presence of a QP decomposition, we consider a more general effect of observed covariates on the random effect distribution, via an appropriate model for the prior masses. This is the focus for section \ref{FMmod}. This proposal may be extended without increasing the number of involved model parameters to regression models with random coefficients of any dimension. This is a clear advantage we get from the proposed approach. The empirical behaviour of the proposal is evaluated using a large scale simulation study, detailed in section \ref{simulatedata}, while the analysis of a well-knowwn benchmark dataset is discussed in section \ref{empdata}. 
Last, section \ref{conclusions} summarizes the paper, and proposes future research agenda.

\section{Random effect models for longitudinal data}\label{longdata}
A longitudinal study is based on the observation of a two-stage sample

\[
\left\{y_{it}, {\bf x}_{it}\right\} \quad i=1,\dots,n,\ t=1,\dots,T,
\]

\noindent where the same units $i=1,\dots, n$ are followed in time, and observations are recorded at a number of (usually common and pre-specified) time occasions, indexed by $t \in \left\{1,\dots,T\right\}$. Here, $y_{it}$ represents the observed value of the response $Y_{it}$ for the $i$-th individual at the $t$-th occasion, and ${\bf x}_{it}$ denotes the corresponding $p$-dimensional design vector that account for \emph{observed} heterogeneity. We will consider equally spaced time occasions for sake of simplicity, but the following arguments may apply to more general designs also. 

According to \cite{andr:etal:2013}, a longitudinal study may help measuring changes at the individual level, separating so-called \emph{age} and \emph{cohort} effects; it allows to control for omitted variable bias and assess potential causal relations. A further by-product is that, by means of a longitudinal study, we may obtain a larger (and often more informative) sample. A major issues in longitudinal studies is that measurements recorded on the same unit are likely dependent, and this dependence should be considered to define a proper modelling approach. For this purpose, it could be useful to fix differences between different sources of dependence; according to the taxonomy introduced by \cite{heck:1981}, we have an \emph{apparent} contagion when analysed individuals are characterized by a constant, individual-specific, unobserved propensity to the event of interest. Therefore, the repeated measurements from the same individual are dependent since they share some common, latent features. On the other hand, so-called \emph{true} contagion describes those situations where past outcomes directly influence (without any mediation) current and future outcomes, causing changes over time in the corresponding distribution. In the following, we will focus on apparent contagion and (relatively) short longitudinal studies.

In a Gaussian context, we may specify an appropriate covariance structure to describe this form of within individual dependence; however, this may not be feasible in general non-Gaussian contexts, where multivariate distributions are seldom available. In such cases, therefore, it is customary to insert, in the linear predictor, individual-specific time-constant random effects, say ${\bf u}_i$, to describe such (constant) within-individual dependence, due to individual-specific \emph{unobserved} heterogeneity.

The modelling approach could be defined by assuming a parametric (conditional) distribution for the observed response, e.g. a member of the exponential family
\[
Y_{it} \mid {\bf u}_{i},\mathbf{x}_{it}\sim {\rm EF}(\theta_{it}),
\]
$i=1,\dots,n$, $t=1,\dots,T$, indexed by a canonical parameter $\theta_{it}$ that is linked to the expected value via the \emph{mean} function
\[
{\rm E}\left(Y_{it} \mid {\bf u}_{i},\mathbf{x}_{it}\right)=\mu_{it}=m(\theta_{it}).
\]
A standard random effect model describes the conditional expectation, where $g(\cdot)$ denotes the so-called \emph{link} function
\[
g(\mu_{it}) = {\bf x}_{it}^{\prime}\bbeta + {\bf w}_{it}^{\prime} {\bf u}_{i}.
\]
In general, ${\bf w}_{it} \subseteq {\bf x}_{it}$; however, we will discuss only the case ${\bf w}_{it}=1$ which defines a random intercept model. The marginal longitudinal model is completed by specifying an appropriate distribution for the random effects 

\[
U_{i} \sim f_{U}(\cdot \mid \bphi). 
\]

As noted before, the random effects are meant to represent individual-specific \emph{unobserved} heterogeneity, while \emph{observed} heterogeneity is summarized by the design vector ${\bf x}_{it}$. The resulting structure of dependence is quite simple, at least on the linear predictor scale. Further, more general, dependence structures have been considered to model specific empirical situations. For example, \cite{cele:etal:2005}, \cite{ng:mcla:2005} discuss genetic studies with technical replicates and multivariate measurements for the same gene. The  developments in such a three-way data context can be dated back at least to \cite{basf:mcla:1985}, see also \cite{verm:2007} and \cite{viro:2011} for more recent proposals in the model-based clustering context. For an interesting review of more complex proposals, the interested reader is referred to \cite{mart:verm:2013}, who give a unifying review of mixtures of \emph{Linear Mixed Models} (LMM) using a general framework, based on Structural Equation Models. 

Assuming independence between the individuals and the repeated measurements from the same individual, in the last case conditional on the individual-specific latent effects, the marginal likelihood is obtained integrating the $u_i$s out

\begin{eqnarray}
L \left(\cdot \right) &=& \prod\limits_{i = 1}^{n} \left\{{\int\limits_{\mathcal{U}} {f_{Y \mid X,U}\left( {\bf y}_{i} \mid {{\bf x}_{i} , u_{i}} \right)}}
f_{X \mid U}(u_i \mid {\bf x}_{i}) \mathrm{d} u_i\right\} \nonumber \\
&=&  
\prod\limits_{i = 1}^{n}
\left\{{\int\limits_{\mathcal{U}} \left[\prod\limits_{t = 1}^{T}
{f_{Y \mid U, X}\left( y_{it} \mid {{\bf x}_{it} , u_{i}} \right)}\right] }
f_{U \mid X}(u_i \mid {\bf x}_{i}) \mathrm{d} u_i\right\} \label{randintlik}
\end{eqnarray}
where ${\bf y}_{i}=\left(y_{i1},\dots,y_{iT}\right)^{\prime}$ and ${\bf x}_{i}=\left[{\bf x}_{i1}, \dots, {\bf x}_{iT}\right]^{\prime}$. The key point in the marginal likelihood expression is that the integral is defined with respect to the density $f_{U \mid X}(u_i \mid {\bf x}_{i})$ which may be, and in general is, different from the distribution assumed for the random effects, that is often, in the literature, parametric and independent of ${\bf x}_{i}$. We will focus on this standard assumption with more details in the following paragraph.

\section{Handling endogeneity: auxiliary regression}\label{aux}
As we may notice, the likelihood in eq. (\ref{randintlik}) is derived considering the $u_i$s as nuisance parameters and integrating them out. The integral is defined with respect to the conditional density $f_{U \mid X}(u_{i} \mid {\bf x}_{i})$ of the random effects given the observed heterogeneity. If we assume that the individual-specific random effects approximate the impact of unobserved time-constant covariates (and therefore represent so called \emph{unobserved} heterogeneity), this is motivated by the potential dependence (in the linear case correlation) between the observed and the unobserved covariates. However, we have no information on this conditional density, as our basic hypotheses entail the marginal distribution of the random effects.

A first, naive, approach may be defined by assuming $f_{U \mid X}(u_{i} \mid {\bf x}_{i})=f_{U}(u_{i})$. This leads to the so-called \emph{random effect} estimator (the GLS estimator in the Gaussian case); in the linear case, the assumption is that of strict \emph{exogeneity} of ${\bf x}_{i}$, that is ${\rm E}(U_{i}\mid {\bf x}_{i})={\rm E}(U_{i})=0$, $i=1,\dots,n$, motivated by ensuring \emph{identifiability} of the common intercept. A further, more refined, option is resuming to a conditional ML approach, that does not need to specify neither $f_{U \mid X}(u_{i} \mid {\bf x}_{i})$ nor $f_{U \mid X}( u_{i})$. In the (conditionally) Gaussian case with identity link, the conditional maximum likelihood approach leads to the fixed effects estimator obtained by demeaning, see e.g. \cite{hsia:2003}. In the generalized linear models framework, however, the conditional maximum likelihood (aka fixed effects approach) is, however, not fully exploited; in fact, it needs a sufficient statistics for the individual-specific effects to exist, which is not always the case for general link functions. A popular example is the complementary log-log link for (conditionally) binary responses, see e.g. \cite{goet:vans:2008}. See \cite{cham:1980} and \cite{cona:1989} for a discussion of the conditional ML approach in the binary response case. 

The assumption of \emph{exogeneity} (independence in the general case) is often verified by the Hausman test, which is used to choose between a conditional fixed effect-type estimator and a marginal, potentially misspecified random effect estimator. However, the hypothesis which is currently tested is that $f_{U \mid X}(u_{i} \mid {\bf x}_{i})=f_{U}(u_{i})$ or, if we look at the linear mixed effect model, ${\rm E}(U_{i}\mid {\bf x}_{i})={\rm E}(U_{i})=0$. In the case of rejection, this does not necessarily mean that a marginal approach can not be taken; rather, the assumption of independence is rejected  by the observed data and, therefore, the model is potentially misspecified. The source of misspecification is the assumption of independence between the observed and the unobserved heterogeneity. 
Therefore, we may still rely on a marginal approach, if we solve the issue of how $f_{U \mid X}(u_{i} \mid {\bf x}_{i})$ should be handled. 

A first attempt to parametrically model this dependence is the \emph{auxiliary} regression approach due to \cite{mund:1978}. To describe this approach, let us consider the likelihood for a simple random intercept model, and do not assume, as it is usual, $f_{U \mid X}(u_{i} \mid {\bf x}_{i})=f_{U}(u_{i})$.

To account for the (linear) dependence of $u_{i}$ on ${\bf x}_{i}$, \cite{mund:1978} suggested to project the random effects onto the space spanned by the ${\bf x}$s, as follows
\begin{equation}\label{auxreg}
u_{i}= u_{i}^{*} + {\rm E}\left(u_{i} \mid {\bf x}_{i}\right)= u_{i}^{*} + \sum_{t}{\bf x}_{it}^{\prime} \bgamma_{t}
\end{equation}
where the first term ($u_{i}^{*}$) is a residual of the regression of $u_{i}$ on ${\bf x}_{i}$ and, therefore, it is linearly free of ${\bf x}_{i}$. This suffices for at least the \emph{weak} exogeneity ${\rm E}\left(u_{i}^{*} \mid {\bf x}_{i}\right)={\rm E}\left(u_{i}^{*}\right)$ condition to hold. In the simplest case, $\bgamma_{t}= \bgamma$, $t=1,\dots,T$ and
\[
{\rm E}\left(u_{i} \mid {\bf x}_{i}\right)=\sum_{t}{\bf x}_{it}^{\prime} \bgamma =T \frac{1}{T} \sum_{t}{\bf x}_{it}^{\prime} \bgamma=\bar {\bf x}_{i}^{\prime} T \bgamma=\bar {\bf x}_{i}^{\prime} \bdelta \, ,
\]
where $\bdelta=T \bgamma$. Following \cite{mund:1978}, for a (conditionally) Gaussian response, we get the following model structure
\[
\left\{
\begin{array}{ll}
{\rm E}\left(Y_{it} \mid {\bf x}_{it}, u_{i}\right) = {\bf x}_{it}^{\prime} \bbeta + u_{i} \\
u_{i} =  u_{i}^{*} + \bar{\bf x}_{i}^{\prime} \bdelta &\\
u_{i}^{*} \sim f_{U^{*}}(\cdot \mid \phi) & \\
\end{array}
\right.
\]

This parametrization has been extended by \cite{cham:1982}, and the corresponding estimator is usually referred to as the \emph{correlated random effect} estimator. More recently, \cite{neuh:kalb:1998} and \cite{neuh:mccu:2006} have discussed this approach in multilevel regression models, with a focus on \emph{conditioaning} and \emph{partitioning}. The interest in this topic is flowering once again, see e.g. \cite{lamo:etal:2016}, who propose a review and an annotated discussion in the context of regression models for multilevel data. By doing a little algebra, we obtain the so called \emph{QP} decomposition, \cite{kris:2005}
\begin{eqnarray*}
{\rm E}\left(Y_{it} \mid {\bf x}_{i}, u_{i}\right) & = & {\bf x}_{it}^{\prime} \bbeta + u_{i} = 
{\bf x}_{it}^{\prime} \bbeta + \bar{\bf x}_{i}^{\prime} \bdelta + u_{i}^{*} = \\
& = & \left({\bf x}_{it}-\bar{\bf x}_{i}\right)^{\prime} \bbeta + \bar{\bf x}_{i}^{\prime} \left(\bdelta + \bbeta \right) + u_{i}^{*}
\end{eqnarray*}
where $\bbeta$ and $(\bdelta + \bbeta)$ represent the \emph{within} and the \emph{between} effects, respectively. 


In the linear (Gaussian) mixed model, it has been proved that estimates obtained by the random effect approach with a QP decomposition are \emph{numerically} equal to those obtained by the fixed effect estimator, for balanced panels. This is obviously true when we consider the parameters associated to time-varying covariates, while the effects for time-invariant covariates are readily (and consistently) estimated in the random effect case, while an additional (between) model should be considered in the fixed effect case to derive estimates for these model parameters. In the nonlinear case, as well as in the case of unbalanced panel, this equality does not hold anymore; results in \cite{neuh:kalb:1998} and \cite{neuh:mccu:2006} suggest that, with increasing $n$ the two are nearly equal, even if some results in \cite{brum:etal:2010} and \cite{brum:etal:2013} seem to cast some doubts in quite extreme conditions. Using this parametrization, we account for a particular form of linear dependence between the observed covariates and the random effects. Such a solution seems to be satisfactory as it leads to parameter estimates that, in the balanced case, \emph{coincide} with those obtained via the fixed-effect approach. However, there is no guarantee that the same approach is equally efficient in the case of non-linear models and/or non Gaussian responses. Finally it is important to say that, when additional information on covariates' endogeneity is available, we may employ an instrumental variables (IV) approach as proposed by \cite{haus:tayl:1981}. In the next section we provide an alternative solution, by adopting a semi-parametric approach that helps account for general forms of linear and non-linear dependencies.


\section{The choice of the mixing distribution}\label{mixing}
As it can be easily noticed by looking at the likelihood in equation (\ref{randintlik}), the individual-specific latent effects are treated as nuisance parameters and integrated out. Apart from the specific case of (conditionally) Gaussian response and Gaussian random effects, and a few more cases (based on conjugate distributions for the random effects), the integral for the likelihood function does not have a closed form and needs to be approximated. A huge literature focused on the choice of an appropriate form for the random effect distribution, as well as on (mainly numerical) approaches to solve the integral in the likelihood equation. For example, parametric (continuous) mixing distributions have been proposed in a Monte Carlo Maximum Likelihood context by \cite{mccu:1994} and \cite{jank:boot:2003}, the latter comparing Monte Carlo and simulated maximum likelihood approaches. Simulated ML has been proposed by \cite{munk:triv:1999}. Approximate ML approaches based on standard Gaussian quadrature have been proposed by \cite{pier:sand:1975}, \cite{ande:aitk:1985}, while, more recently, adaptive Gaussian quadrature has received some attention, see e.g. \cite{liu:pier:1994}, \cite{pine:bate:1995} and \cite{rabe:skro:2002}. Spherical quadrature has been dealt with by \cite{rabe:skro:2005}, pseudo adaptive quadrature has been introduced in highly parametrized joint models for a longitudinal outcome and a time to event, see \cite{rizo:2012}, while further options may also be available, see e.g. \cite{lang:king:2013}. Simulation-based comparisons have been discussed by \cite{lesa:spie:2001} and, more recently in a mixed HMM framework, by \cite{mari:alfo:2016}. Semi-parametric approaches based on a discrete estimate for the mixing distribution have been introduced, as a counterpart to Gaussian quadrature to handle \emph{overdispersion} in \cite{hind:wood:1987}, or \emph{unobserved heterogeneity} in \cite{foll:lamb:1989}, \cite{diet:boeh:1994}. Early applications can be found in marketing, see e.g. \cite{wede:desa:1995}, biostatistics, see \cite{wang:etal:1996}, econometrics, \cite{deb:triv:1997}, and machine learning \cite{jaco:etal:1991}. However, \cite{aitk:1996} and \cite{aitk:1999} seem to be the first attempts to highlight the connection between mixed effect models and finite mixtures, via the theory of nonparametric maximum likelihood estimation (NPMLe) of the mixing distribution, introduced by \cite{kief:wolf:1956}, revisited by \cite{sima:1976}, \cite{lair:1978}, \cite{boeh:1982} and fully developed by \cite{lind:1983a,lind:1983b}. This approach will be further detailed in the following section. While the discrete semi-parametric approach has several advantages over the parametric ones, it has been criticized as a discrete distribution for the random effects may be unrealistic, see e.g. \cite{neuh:mccu:2011a, neuh:mccu:2011b}. For this and related reasons, several non- or semi-parametric extensions have been discussed in the literature; for example, \cite{davi:gall:1993} propose an unrestricted, smooth, mixing distribution, \cite{magd:zege:1996} discuss a  smooth version of the NPMLe, similar to the mixtures of Gaussian densities proposed by \cite{verb:lesa:1996, verb:lesa:1997}. Among others, further proposals have been introduced by \cite{ghid:etal:2005}, who use penalized B-splines and by \cite{vivi:etal:2013} who discuss a \emph{``nonparanormal''} approach which, however, poses some estimation problems that have stopped further developments. All these approaches trade computational complexity for a more flexible and smooth random effect distribution. Several results seem to support low sensitivity of parameter estimates to assumptions regarding the mixing distribution. According to some authors, however, see e.g. \cite{wang:tayl:2001}, \cite{song:etal:2002}, \cite{tsia:davi:2004}, \cite{rizo:etal:2008}, misspecification of the random effect distribution may have great impact on some parameter estimates, especially when we face longitudinal studies with a low number of measurement occasions per individual. This issue is investigated with great details by \cite{neuh:mccu:2011a, neuh:mccu:2011b}. However, there is not general agreement on such results or, better, results are highly dependent on simulation schemes and parameters we are interested in. Just to give an example, simulations performed by \cite{liti:etal:2008} support potentially high impact on some parameter estimates, in particular on individual-specific, time-constant, parameters. Rather than assuming a formal, and potentially erroneous, position, we notice that, as usual, a sensitivity analysis is mandatory, and semi-parametric approaches should be considered together with parametric ones.
Based on the results by \cite{sima:1976}, \cite{lair:1978}, \cite{boeh:1982} and \cite{lind:1983a,lind:1983b}, as long as the likelihood in equation (\ref{randintlik}) is bounded, it is maximized wrt ${F}_{U}(\cdot)$ by at least a discrete distribution with at most $K \leq n_{d}$ support points (NPMLe), where $n_{d}$ represents the number of distinct profiles (response, design vector) in the analysed sample. In particular, the results discussed by \cite{lind:1983a,lind:1983b} are of prominent importance as it is shown that maximizing the likelihood in eq. (\ref{randintlik}) reduces to the standard problem of maximizing a concave function over a convex set, see also \cite{mcla:peel:2000}. Using the QP decomposition in eq. (\ref{auxreg}) the likelihood function can be written
\begin{eqnarray}
L\left(\cdot \right) & = & \prod\limits_{i = 1}^{n}
{\int\limits_{\mathcal{U}} \left[\prod\limits_{t = 1}^{T}
{f_{Y \mid U, X}\left( y_{it} \mid {{\bf x}_{it} , u_{i}} \right)}\right] }
f_{U}(u_i \mid {\bf x}_{i}) \mathrm{d}u_i = \nonumber \\
& = & \prod\limits_{i = 1}^{n}
{\int\limits_{\mathcal{U}} \left[\prod\limits_{t = 1}^{T}
{f_{Y \mid U^{*}, X}\left( y_{it} \mid {{\bf x}_{i}, u^{*}_{i}} \right)}\right] }
f_{U^{*}}(u^{*}_i ) \mathrm{d}u^{*}_i
\label{QPlik}
\end{eqnarray}
According to the theory of NPMLe, the integral in the likelihood can be approximated by a finite sum
\begin{equation}
L\left(\cdot \right) \simeq \prod\limits_{i = 1}^{n}
\left\{{\sum\limits_{k = 1}^K  \left[\prod\limits_{t =
1}^{T} {f_{Y \mid U^{*}, X}\left( {y_{it} \mid {{\bf x}_{i} , \zeta_k}}
\right)}\right]}\pi _k \right\} = \prod\limits_{i = 1}^{n}
\sum\limits_{k = 1}^K f\left({\bf y}_{i} \mid {\bf x}_{i} , \zeta_k\right) \pi _k
\label{approximatelik}
\end{equation}
where $u_i^{*} \sim \sum_{k=1}^{K} \pi_{k} \mathds{1}(u_{i}^{*}=\zeta_{k})$, and $\pi_k=\Pr({\bf U}_{i}=\bzeta_{k})$, $k=1,\dots,K$. 
Model identifiability is discussed in \cite{wang:etal:1996}, where full rank condition for the covariates matrix are given, while more stringent conditions are provided by \cite{foll:lamb:1991}, \cite{henn:2000}, \cite{grun:leis:2008}. The last two formally discuss methods for bounding $K$ from above to ensure model identifiability, see also \cite{fruh:2006} for a thoughtful review of the topic. \cite{ders:1986} describes the details of the EM algorithm for ML estimation. We summarize its structure below.

\subsection{ML estimation via the EM algorithm}
The estimation problem may be posed as an \emph{incomplete} data problem, where the \emph{missing} information is the component indicator $z_{ik}=\mathds{1}(u_{i}^{*}=\zeta_{k})$, $i=1,\dots,n$, $k=1,\dots,K$. For fixed (and known) $K$, let us denote the unobservable membership vector and the \emph{global} model parameter vector by $\textbf{z}_i = (z_{i1},z_{i2},\dots,z_{iK})$, and $\bxi$. In this framework, the EM algorithm arises quite naturally as the appropriate choice for ML estimation. The starting point is the so called \emph{complete data} log-likelihood function
\[
\ell_{c}\left(\bxi, \bpi\right) = \sum\limits_{i = 1}^{n}\sum\limits_{k
= 1}^K z_{ik}\left[\log(\pi_k )+\log f({\bf y}_{i} \mid {\bf x}_{i}, \zeta_{k})\right]
\]
which should be our objective function if we would have been able to observe ${\bf z}$. However, the component indicator is missing and, therefore, the EM algorithm is based on a \emph{surrogate} objective function based on \emph{complete data}; this is the posterior expectation of the \emph{complete data}, based on current parameter estimates and conditional on observed data, $Q\left(\bxi, \bpi \mid \bxi^{(r)}, \bpi^{(r)} \right)$; instead of the log-likelihood function the EM algorithm is based on computing this expectation (E step) and maximizing it with respect to model parameters. Its structure can be briefly summarized as follows:

\begin{itemize}
\item \textbf{E-step}
\begin{eqnarray*}
Q\left(\bxi, \bpi \mid \bxi^{(r)}, \bpi^{(r)} \right) = E_{\bxi^{(r)},\bpi^{(r)}}\left\{\ell_c\left(\bxi, \bpi \right) \mid \textbf{y}\right\} = \\
= \sum\limits_{i = 1}^{n}\sum\limits_{k
= 1}^K \tau_{ik}^{(r+1)}\left[\log(\pi_k )+\log f({\bf y}_{i} \mid {\bf x}_{i}, \bzeta_{k})\right]
\end{eqnarray*}
where $\tau_{ik}^{(r+1)}=\Pr\left(z_{ik}=1 \mid {\bf y}_{i}, \bxi^{(r)}, \bpi^{(r)}\right)$.
\item \textbf{M-step}
To obtain updated parameter estimates, the $Q\left(\cdot\right)$ function is maximized with respect to $(\bxi, \bpi)$, conditional on weights $\tau_{ik}^{(r+1)}$.
\end{itemize}
\noindent Model parameter estimates $\bxi^{(r+1)}$ are the solutions to the following likelihood equations 
\begin{eqnarray}
\frac{\partial Q\left(\bxi, \bpi \mid \bxi^{(r)}, \bpi^{(r)} \right) }{\partial \bxi}=\sum\limits_{i=1}^{n}
\sum\limits_{k=1}^{K} \tau_{ik}^{(r+1)}\frac{\partial} {\partial
\bxi} \log f({\bf y}_{i} \mid {\bf x}_{i}, \bzeta_{k})= 0 \quad . \label{dQzeta}
\end{eqnarray}
The prior weight (mass) estimates $\bpi^{(r+1)}$ come from defining the Lagrange function under the constrain $\sum_{k} \pi_{k}=1$, $\mathcal{L}\left(\bxi, \bpi \mid \bxi^{(r)}, \bpi^{(r)} \right)$ and solving
\begin{eqnarray}
\frac{\partial \mathcal{L}\left(\bxi, \bpi \mid \bxi^{(r)}, \bpi^{(r)} \right)}{\partial{\pi_{l}}}&=& \sum\limits_{i =1}^{n}\left\{\tau_{il}^{(r+1)}-\pi_{l}\right\}=0 \quad . \label{dQpi}
\end{eqnarray}

Once the algorithm has reached a solution for a given number of locations $K$, possibly based on a set of several starting points, we proceed by increasing this number and estimate corresponding model parameters. Thus, the issue raises on how to choose between these solutions, as it is well known that the usual limiting distribution for the LRT does not apply in this context, see e.g. \cite{mcla:peel:2000}. One of the first solutions was proposed by \cite{mcla:1987} and is based on bootstrapping the LRT statistic; probably its use has been severely limited by the computational resources that were available at the end of the 80s. Usually, penalized likelihood criteria such as BIC, \cite{schw:1978}, AIC, \cite{akai:1973}, CAIC, \cite{bozd:1987}, ICL, \cite{bier:etal:2000} are employed in order to choose a parsimonious yet appropriate solution. Related asymptotic results are discussed in \cite{lero:1992} and \cite{keri:2000}, while simulation based results can be found in \cite{karl:meli:2007}.
Despite the similarities between the likelihood in eq. (\ref{approximatelik}) and the likelihood of a finite mixture model, we should notice that NPMLe requires the likelihood to be maximized with increasing $K$, without considering any penalization term. On the contrary, penalized likelihood criteria are usually employed in finite mixture models, where the number of components is a key issue (eg  in a model-based clustering context). In the present context, the finite mixture gives just an approximation to a possibly continuous mixing distribution, without a \emph{true} number of components. In other terms, here the mixing distribution is only a \emph{nuisance} component of the model which must be accounted for in order to estimate the parameters of interest. 

The EM algorithm does not provide an estimate of the observed information matrix, as it uses the \emph{complete data} information matrix, the question raises on how to estimate the standard errors for model parameter estimates. Two main approaches have been developed in the last few years: the first relies on (non)parametric bootstrapping, see e.g. \cite{hunt:etal:2007}, \cite{mcla:kris:2008}. The second main approach to standard error estimation is based on using the observed/sandwich information, by adopting the \cite{loui:1982} or the \cite{oake:1999} formulas; \cite{frie:kaue:2000} discuss a simulation study comparing the performance of observed vs sandwich information matrices in a discrete (finite mixture) random intercept generalized linear model. Similar results are detailed by \cite{bold:magn:2008} and \cite{magn:magn:2019}. The observed information can also be approximated by calculating second order partial derivatives and/or the gradient function of the log-likelihood by numerical differentiation, see e.g. \cite{jams:jenn:2000} for a discussion of several issues concerning such an approach. 

A further reason to adopt such a semi-parametric approach is that it leads to a quite straightforward and general solution to \emph{endogeneity} of the design vector. This issue will be developed in the next paragraph.

\section{A solution based on finite mixtures}\label{FMmod}
As we have observed, by using the QP decomposition, we account for a specific form of linear dependence of the random effects on the observed design vector. While this is not of general interest here, when we move from the random intercept to the general random coefficient model, this approach leads to growing dimension of the parameter vector to account for potential dependence between unobserved and unobserved covariates. So, we may wonder whether a different approach can be introduced to account for general dependencies, possibly with a constant dimension for the resulting model. The semi-parametric approach based on finite mixtures gives such an opportunity.
By approximating the unknown random effect distribution via a discrete distribution estimated on the observed data we prevent potential bias in parameter estimates deriving from possibly wrong parametric assumptions on this distribution. However, a more substantial bias can arise from a  (wrong) assumption of exogeneity of the observed covariates. A finite mixture based model can help us to model the dependence betwen the observed covariates and the random effects in a more general and effective way. Until now, the conditioning has been handled by reparametrizing the values for the random effects, but the finite mixture approach allows for the parametrization of the distribution of the random effects via an explicit model for the \emph{prior weights}

\begin{eqnarray*}
L\left(\cdot \right) & \simeq & \prod\limits_{i = 1}^{n}
\left\{{\sum\limits_{k = 1}^K \pi_{ik} \left[\prod\limits_{t =
1}^{T} {f_{Y \mid X, Z}\left( {y_{it} \mid {{\bf x}_{it} , \bzeta_k}}
\right)}\right] } \right\} \\
\pi_{ik}& = & \pi_{k}\left(\bar{\bf x}_{i}\right)
\end{eqnarray*}
as in \emph{concomitant variable} models, \cite{dayt:macr:1988}. A popular example is the multinomial logit model
\[
\pi_{il}=\frac{\exp\left(\bar{\bf x}_{i}^{\prime}\bgamma_{l}\right)}{\sum_k \exp\left(\bar{\bf x}_{i}^{\prime}\bgamma_{k}\right)}, \quad l=1,\dots,K.
\]
where one vector of parameters $\bgamma_{k}$, $k=1,\dots,K$, is usually set to $\bf0$ to ensure identifiability. As it can be easily noticed, the impact of the observed covariates on the random effects is accounted for by the model for the prior weights. Such a parameterization leads to a simpler interpretation, and allows to take into account more general forms of dependence between the random effects and the observed covariates. Further, it can be extended to higher dimensional, individual-.specific, random coefficient without the need to increase model complexity in terms of estimated parameters in the model for the prior weights. It is worth noticing that we may use this parametrization any time some of the adopted covariates (say observed explanatory variables, autoregressive components, indicators of missing information, etc.) can be considered as dependent on the random effects. As usual, we are not be able to distinguish whether the effect of time-invariant covariates are direct on the observed response or mediated through the individual-specific random effects. That is, time invariant covariates can only be associated to between effects. Therefore, we have to decide whether such covariates should be included in the linear predictor or in the model for the prior weights, as these two effects can not be properly distinguished. Depending on the kind of parametrization for the prior weights (logit, ordinal logit, etc.), the proposed approach may be more/less parsimonious than the conditioning handled through shift in the random effect locations, through the auxiliary regression in eq. (\ref{auxreg}). In the multinomial logit parameterization, eq. (\ref{dQpi}) modifies to:
\begin{equation}
\frac{\partial \mathcal{L}\left(\bxi, \bgamma \mid \bxi^{(r)}, \bgamma^{(r)} \right)}{\partial{\bgamma_{l}}} = \sum\limits_{i =1}^{n}\tau_{il}^{(r+1)}\left(1-\pi_{il}\right)\bar{\bf x}_{i}=0 \quad . \label{dQgamma}
\end{equation}

The proposed model shows interesting connections with modelling approaches introduced in several, quite different, contexts. Examples are the \emph{Latent Dropout Classes model} introduced by \cite{roy:2003}, and extended by \cite{roy:dani:2008}, the heterogeneous Hidden Markov models  described by \cite{maru:rocc:2012}, or by \cite{mari:alfo:2015} in the quantile regression context. Further examples are the models proposed, independently, by \cite{aitk:alfo:1998} and \cite{wool:2005} to solve the so called \emph{initial conditions} problem, and by \cite{alfo:maru:2009} to account for potentially non-ignorable missingness.
These models have never been considered, at least to our knowldge, as specific cases of a more general model for handling endogeneity.

\section{Simulation Study}\label{simulatedata}
This section presents a large-scale simulation experiment designed to study the empirical behaviour of the proposed modelling approach, and compare it with potential competitors. For this purpose, we consider the following simulation scheme. $B=250$ samples have been drawn from either a (conditional) Gaussian or a Bernoulli distribution, with

\[
Y_{it} \mid {\bf x}_{it}, u_{i} \sim \left\{
\begin{array}{l}
{\rm Bin}(1, \mu_{it}) \\
{\rm N}(\mu_{it}, \sigma_{e}^{2})
\end{array}\right.
\]
The conditional expectation is described by the following linear (on the link function scale) mixed effect model
\[
g \left( \mu_{it} \right) = \beta_0 + \beta_1 x_{it} + u_{i}
\]
where regression coefficients for each sample have been uniformly drawn from the following ranges $\beta_0 \in \left(-0.6, -0.2\right)$ and $\beta_1 \in \left(0.25, 0.75\right)$. As far as the covariates values are concerned, they have been drawn from a multivariate ($T$-dimensional) Gaussian density, that is ${\bf x}_{i} \sim {\rm MVN}_{T}(\bmu_{x}, {\bf \Sigma_{x})}$, where terms have unit variances and a constant correlation. The link function $g(\cdot)$ is the \emph{identity} in the Gaussian case, and the \emph{probit} in the Bernoulli case, in order to consider also non \emph{canonical} models.

For each of the two scenarios (A: Gaussian, B: Bernoulli) we have considered 3 potential sub-scenarios:
\begin{itemize}
\item[Scenario 1] Gaussian individual-specific random effects with constant correlation $\rho=cor \left(u_{i}, x_{it} \right)$, $\forall t=1,\dots,T$; the value of the correlation coefficient has been uniformly drawn from three possible intervals corresponding to low, medium and high dependence, respectively:
\begin{itemize}
\item[1.1] $\rho \in (0, 0.2)$
\item[1.2] $\rho \in (0.2, 0.5)$
\item[1.3] $\rho \in (0.5, 0.8)$
\end{itemize}
\item[Scenario 2]
$u_{i} = \exp(\gamma_{0} + \gamma_{1}\bar{x}_{i}) + \varepsilon_{i}$, where $\varepsilon_{i} \sim {\rm N}(0,1)$
\item[Scenario 3] $u_{i} \sim \sum_{k=1}^{K} \pi_{k}(\bar{x}_{i}) \mathds{1}(u_{i}=\zeta_{k})$, $K=3$, and
\[
\log\frac{\pi_{k}(\bar{x}_{i})}{\pi_{K}(\bar{x}_{i})}=\phi_{0k}+\phi_{1k}\bar{x}_{i}, \quad k=1,\dots,(K-1)
\]
with $\bzeta=(-2,0,1)$, $\bphi_{0}=(0, 0.5, -3.5)$, $\bphi_{1}=(0, -3.5, 3)$.
\end{itemize}
For the Bernoulli case only, we have also considered the following additional scenario:
\begin{itemize}
\item[Scenario 4]  
$u_{i} = \gamma_{0}+\gamma_{1}\left(\max_{t}{x}_{it}\right) + \varepsilon_{i}$, where $\varepsilon_{i} \sim {\rm N}(0,1)$, with the purpose of comparing results with those obtained by \cite{brum:etal:2010}
\end{itemize}

For each sample, we have considered the present proposal (COV in the following Tables) a semi-parametric random effect model based on a finite mixture (FM),  with a QP decomposition (FMQP), a parametric random effect model (Par), with a QP decomposition (ParQP). In the case of binary responses, we have also considered the fixed effect estimator (FE), and the bias-corrected fixed effect estimator (FEbc), see \cite{hahn:2004}, from the R package \texttt{bife}, see \cite{stam:2016}. The FE estimator has not been taken into account in the Gaussian case, as it can be shown to be equal, in the present balanced study case, to the estimator obtained by ParQP, see \cite{mund:1978} or \cite{balt:2009}. For all the models based on finite mixtures, we used three different methods to select the number of components: likelihood increment (with threshold $\epsilon=10^{-07}*npar(K)$ between two subsequent values $K$ and $(K+1)$), AIC \cite{akai:1973}, or BIC \cite{schw:1978}, where $npar(K)$ represents the number of parameters for a model with $K$ components. We are not saying that the likelihood for a mixture of Gaussians can be maximized with respect to $K$, as this is well know to be unbounded, see \cite{day:1969}, \cite{hath:1985} and \cite{boeh:1995}, among others. Rather, we think the critierion of a small threshold could be used, at least in the case of a continuous true \emph{mixing} distribution, as a starting solution for further enhancements.

We have considered different values for the sample size $n \in \left\{100, 250, 500\right\}$ and the number of time occasions for each individual $T=5, 10$.
We report, for each of the 10 different scenarios we considered, the average bias ($bias$), the average squared error ($ASE$), calculated as 
\[
ASE(\beta_j)= \frac{1}{B}\sum_{b=1}^{B}\left(\widehat{\beta}_{jb} -\beta_j\right)^2
\]
and the sample standard deviation ($sd$) for $\bbeta$ estimates.

Let us start from the scenarios regarding a (conditional) Gaussian response.
\begin{center}
\textbf{Tables \ref{table1}--\ref{table5} about here}
\end{center}

As it can be evinced by looking at tables \ref{table1}--\ref{table3}, the \emph{standard} (misspecified) random effect estimator leads to substantially biased estimates, and the bias is increasing with correlation between the omitted (represented by the random effects) and the observed covariates. The bias is specifically high in the presence of \emph{short} individual sequences ($T=5$) for any value of the sample size $n$. When we look at the QP decomposition, the estimates obtained by fitting a parametric mixed model behave slightly better than those from the corresponding semi-parametric model, based on finite mixtures. When we look at our proposal, we may notice a good behaviour which is particularly good when the number of components of the finite mixture is chosen by the maximized likelihood value only (with a minimum threshold), suggesting that both AIC and BIC tend to penalize too much. We must notice that, in settings A1--A3, the ParQP specification refer to the \emph{true} data generating model while all the remaining are derived from somewhat misspecified models. However, even when we move from the linear dependence case to more complex dependence structure, as those in tables  \ref{table4}--\ref{table5}, the ParQP still remains the reference model even if our proposal tends to have very similar behaviour in terms of both ASE and bias when the number of components is chosen according to the maximized likelihood value. As we have mentioned before, we have not considered the fixed effect estimator, since the resulting estimate is numerically equal, in the case of balanced panels, to that obtained by ParQP.
\begin{sidewaystable}[!ht]
\begin{tabular}{ccc|rrrrrrrrrrr}
n	&	T	&		&	FMLik	&	FMAIC	&	FMBIC	&	CovLik	&	CovAIC	&	CovBIC	&	FMQPLik	&	FMQPAIC	&	FMQPBIC	&	PAR	&	PARQP	\\ \hline \hline
100	&	5	&	ASE	&	0,0039	&	0,0040	&	0,0043	&	0,0034	&	0,0036	&	0,0039	&	0,0033	&	0,0033	&	0,0033	&	0,0039	&	0,0033	\\
	&		&	Bias	&	0,0266	&	0,0288	&	0,0316	&	0,0049	&	0,0129	&	0,0132	&	0,0043	&	0,0043	&	0,0043	&	0,0275	&	0,0043	\\
	&		&	sd	&	0,0563	&	0,0564	&	0,0577	&	0,0581	&	0,0590	&	0,0610	&	0,0576	&	0,0576	&	0,0576	&	0,0560	&	0,0576	\\ \hline
100	&	10	&		&	0,0015	&	0,0015	&	0,0016	&	0,0011	&	0,0012	&	0,0012	&	0,0011	&	0,0011	&	0,0011	&	0,0014	&	0,0011	\\
	&		&		&	0,0172	&	0,0186	&	0,0205	&	0,0028	&	0,0060	&	0,0070	&	0,0027	&	0,0027	&	0,0027	&	0,0176	&	0,0027	\\
	&		&		&	0,0340	&	0,0346	&	0,0347	&	0,0333	&	0,0340	&	0,0342	&	0,0335	&	0,0335	&	0,0335	&	0,0337	&	0,0335	\\ \hline
250	&	5	&		&	0,0017	&	0,0018	&	0,0019	&	0,0011	&	0,0011	&	0,0011	&	0,0011	&	0,0011	&	0,0011	&	0,0017	&	0,0011	\\
	&		&		&	0,0235	&	0,0243	&	0,0260	&	-0,0011	&	0,0014	&	0,0018	&	-0,0012	&	-0,0012	&	-0,0012	&	0,0242	&	-0,0012	\\
	&		&		&	0,0341	&	0,0343	&	0,0346	&	0,0327	&	0,0331	&	0,0333	&	0,0327	&	0,0327	&	0,0327	&	0,0341	&	0,0327	\\ \hline
250	&	10	&		&	0,0007	&	0,0008	&	0,0008	&	0,0004	&	0,0004	&	0,0005	&	0,0004	&	0,0004	&	0,0004	&	0,0007	&	0,0004	\\
	&		&		&	0,0145	&	0,0152	&	0,0159	&	0,0009	&	0,0027	&	0,0029	&	0,0008	&	0,0008	&	0,0008	&	0,0146	&	0,0008	\\
	&		&		&	0,0229	&	0,0234	&	0,0235	&	0,0207	&	0,0209	&	0,0213	&	0,0205	&	0,0205	&	0,0205	&	0,0227	&	0,0205	\\ \hline
500	&	5	&		&	0,0012	&	0,0013	&	0,0013	&	0,0007	&	0,0006	&	0,0007	&	0,0007	&	0,0007	&	0,0007	&	0,0012	&	0,0007	\\
	&		&		&	0,0206	&	0,0211	&	0,0222	&	-0,0042	&	-0,0011	&	-0,0013	&	-0,0041	&	-0,0041	&	-0,0041	&	0,0210	&	-0,0041	\\
	&		&		&	0,0283	&	0,0285	&	0,0290	&	0,0257	&	0,0240	&	0,0260	&	0,0253	&	0,0253	&	0,0253	&	0,0281	&	0,0253	\\ \hline
500	&	10	&		&	0,0005	&	0,0005	&	0,0005	&	0,0003	&	0,0002	&	0,0003	&	0,0003	&	0,0003	&	0,0003	&	0,0005	&	0,0003	\\
	&		&		&	0,0126	&	0,0129	&	0,0135	&	-0,0013	&	0,0001	&	0,0001	&	-0,0013	&	-0,0013	&	-0,0013	&	0,0126	&	-0,0013	\\
	&		&		&	0,0176	&	0,0178	&	0,0180	&	0,0158	&	0,0130	&	0,0160	&	0,0158	&	0,0158	&	0,0158	&	0,0176	&	0,0158	\\ \hline \hline
\end{tabular}
\caption{Simulation study, scenario 1.1, Gaussian responses and Gaussian random effects, constant correlation $\rho={\rm cor}\left(u_{i}, x_{it}\right) \in \left(0, 0.2 \right]$. ASE, bias and standard deviation of parameter estimates}\label{table1}
\end{sidewaystable}

\begin{sidewaystable}[!ht]
\begin{tabular}{ccc|rrrrrrrrrrr}
n	&	T	&		&	FMLik	&	FMAIC	&	FMBIC	&	CovLik	&	CovAIC	&	CovBIC	&	FMQPLik	&	FMQPAIC	&	FMQPBIC	&	PAR	&	PARQP	\\ \hline \hline
100	&	5	&	ASE	&	0,0188	&	0,0213	&	0,0237	&	0,0039	&	0,0067	&	0,0067	&	0,0040	&	0,0040	&	0,0040	&	0,0196	&	0,0040	\\
	&		&	Bias	&	0,1179	&	0,1264	&	0,1347	&	-0,0020	&	0,0357	&	0,0360	&	-0,0011	&	-0,0011	&	-0,0011	&	0,1216	&	-0,0011	\\
	&		&	sd	&	0,0701	&	0,0732	&	0,0746	&	0,0626	&	0,0736	&	0,0738	&	0,0632	&	0,0632	&	0,0632	&	0,0691	&	0,0632	\\ \hline
100	&	10	&		&	0,0066	&	0,0074	&	0,0088	&	0,0018	&	0,0025	&	0,0025	&	0,0019	&	0,0019	&	0,0019	&	0,0071	&	0,0019	\\
	&		&		&	0,0644	&	0,0695	&	0,0769	&	-0,0009	&	0,0136	&	0,0138	&	-0,0017	&	-0,0017	&	-0,0017	&	0,0681	&	-0,0017	\\
	&		&		&	0,0494	&	0,0507	&	0,0535	&	0,0428	&	0,0482	&	0,0484	&	0,0431	&	0,0431	&	0,0431	&	0,0500	&	0,0431	\\ \hline
250	&	5	&		&	0,0186	&	0,0197	&	0,0216	&	0,0016	&	0,0025	&	0,0025	&	0,0016	&	0,0016	&	0,0016	&	0,0191	&	0,0016	\\
	&		&		&	0,1210	&	0,1250	&	0,1311	&	0,0005	&	0,0200	&	0,0210	&	0,0011	&	0,0011	&	0,0011	&	0,1229	&	0,0011	\\
	&		&		&	0,0628	&	0,0639	&	0,0666	&	0,0406	&	0,0455	&	0,0459	&	0,0401	&	0,0401	&	0,0401	&	0,0629	&	0,0401	\\ \hline
250	&	10	&		&	0,0058	&	0,0063	&	0,0069	&	0,0007	&	0,0008	&	0,0009	&	0,0006	&	0,0006	&	0,0006	&	0,0060	&	0,0006	\\
	&		&		&	0,0671	&	0,0701	&	0,0738	&	0,0005	&	0,0099	&	0,0101	&	0,0001	&	0,0001	&	0,0001	&	0,0686	&	0,0001	\\
	&		&		&	0,0361	&	0,0374	&	0,0381	&	0,0257	&	0,0260	&	0,0280	&	0,0249	&	0,0249	&	0,0249	&	0,0363	&	0,0249	\\ \hline
500	&	5	&		&	0,0181	&	0,0187	&	0,0197	&	0,0009	&	0,0012	&	0,0012	&	0,0009	&	0,0009	&	0,0009	&	0,0183	&	0,0009	\\
	&		&		&	0,1244	&	0,1267	&	0,1303	&	-0,0005	&	0,0139	&	0,0141	&	-0,0001	&	-0,0001	&	-0,0001	&	0,1254	&	-0,0001	\\
	&		&		&	0,0509	&	0,0514	&	0,0518	&	0,0307	&	0,0316	&	0,0318	&	0,0305	&	0,0305	&	0,0305	&	0,0510	&	0,0305	\\ \hline
500	&	10	&		&	0,0060	&	0,0063	&	0,0068	&	0,0003	&	0,0004	&	0,0005	&	0,0003	&	0,0003	&	0,0003	&	0,0062	&	0,0003	\\
	&		&		&	0,0710	&	0,0726	&	0,0757	&	0,0001	&	0,0075	&	0,0077	&	0,0001	&	0,0001	&	0,0001	&	0,0719	&	0,0001	\\
	&		&		&	0,0317	&	0,0323	&	0,0334	&	0,0180	&	0,0194	&	0,0198	&	0,0178	&	0,0178	&	0,0178	&	0,0316	&	0,0178	\\ \hline \hline
\end{tabular}
\caption{Simulation study, scenario 1.2, Gaussian responses and Gaussian random effects, constant correlation $\rho={\rm cor}\left(u_{i}, x_{it}\right) \in \left(0.2, 0.5 \right]$. ASE, bias and standard deviation of parameter estimates}\label{table2}
\end{sidewaystable}

\begin{sidewaystable}[!ht]
\begin{tabular}{ccc|rrrrrrrrrrr}
n	&	T	&		&	FMLik	&	FMAIC	&	FMBIC	&	CovLik	&	CovAIC	&	CovBIC	&	FMQPLik	&	FMQPAIC	&	FMQPBIC	&	PAR	&	PARQP	\\ \hline \hline
100	&	5	&	ASE	&	0,1194	&	0,1283	&	0,1385	&	0,0087	&	0,0207	&	0,0209	&	0,0083	&	0,0083	&	0,0083	&	0,1243	&	0,0083	\\
	&		&	Bias	&	0,3295	&	0,3423	&	0,3573	&	-0,0052	&	0,0879	&	0,0884	&	-0,0051	&	-0,0051	&	-0,0051	&	0,3368	&	-0,0051	\\
	&		&	sd	&	0,1043	&	0,1055	&	0,1041	&	0,0930	&	0,1139	&	0,1145	&	0,0911	&	0,0911	&	0,0911	&	0,1043	&	0,0911	\\ \hline
100	&	10	&		&	0,1181	&	0,0754	&	0,0850	&	0,0039	&	0,0106	&	0,0107	&	0,0036	&	0,0036	&	0,0036	&	0,0719	&	0,0036	\\
	&		&		&	0,3288	&	0,2550	&	0,2713	&	0,0062	&	0,0664	&	0,0667	&	0,0041	&	0,0041	&	0,0041	&	0,2484	&	0,0041	\\
	&		&		&	0,1000	&	0,1019	&	0,1066	&	0,0622	&	0,0784	&	0,0788	&	0,0599	&	0,0599	&	0,0599	&	0,1008	&	0,0599	\\ \hline
250	&	5	&		&	0,1230	&	0,1275	&	0,1341	&	0,0029	&	0,0094	&	0,0096	&	0,0029	&	0,0029	&	0,0029	&	0,1251	&	0,0029	\\
	&		&		&	0,3362	&	0,3426	&	0,3524	&	0,0001	&	0,0693	&	0,0699	&	-0,0002	&	-0,0002	&	-0,0002	&	0,3394	&	-0,0002	\\
	&		&		&	0,0996	&	0,1005	&	0,0997	&	0,0543	&	0,0681	&	0,0688	&	0,0540	&	0,0540	&	0,0540	&	0,0995	&	0,0540	\\ \hline
250	&	10	&		&	0,0691	&	0,0727	&	0,0784	&	0,0014	&	0,0035	&	0,0036	&	0,0013	&	0,0013	&	0,0013	&	0,0709	&	0,0013	\\
	&		&		&	0,2437	&	0,2506	&	0,2608	&	0,0014	&	0,0373	&	0,0378	&	-0,0004	&	-0,0004	&	-0,0004	&	0,2475	&	-0,0004	\\
	&		&		&	0,0984	&	0,0997	&	0,1019	&	0,0373	&	0,0462	&	0,0468	&	0,0360	&	0,0360	&	0,0360	&	0,0983	&	0,0360	\\ \hline
500	&	5	&		&	0,1326	&	0,1356	&	0,1401	&	0,0016	&	0,0056	&	0,0057	&	0,0015	&	0,0015	&	0,0015	&	0,1339	&	0,0015	\\
	&		&		&	0,3515	&	0,3557	&	0,3622	&	0,0050	&	0,0546	&	0,0552	&	0,0042	&	0,0042	&	0,0042	&	0,3534	&	0,0042	\\
	&		&		&	0,0951	&	0,0953	&	0,0944	&	0,0397	&	0,0509	&	0,0512	&	0,0387	&	0,0387	&	0,0387	&	0,0950	&	0,0387	\\ \hline
500	&	10	&		&	0,0720	&	0,0739	&	0,0779	&	0,0008	&	0,0018	&	0,0018	&	0,0008	&	0,0008	&	0,0008	&	0,0725	&	0,0008	\\
	&		&		&	0,2498	&	0,2535	&	0,2608	&	0,0019	&	0,0257	&	0,0261	&	0,0018	&	0,0018	&	0,0018	&	0,2509	&	0,0018	\\
	&		&		&	0,0979	&	0,0983	&	0,0992	&	0,0290	&	0,0339	&	0,0341	&	0,0281	&	0,0281	&	0,0281	&	0,0977	&	0,0281	\\ \hline \hline
\end{tabular}
\caption{Simulation study, scenario 1.3, Gaussian responses and Gaussian random effects, constant correlation $\rho={\rm cor}\left(u_{i}, x_{it}\right) \in \left(0.5, 0.8 \right]$. ASE, bias and standard deviation of parameter estimates}\label{table3}
\end{sidewaystable}

\begin{sidewaystable}[!ht]
\begin{tabular}{ccc|rrrrrrrrrrr}
n	&	T	&		&	FMLik	&	FMAIC	&	FMBIC	&	CovLik	&	CovAIC	&	CovBIC	&	FMQPLik	&	FMQPAIC	&	FMQPBIC	&	PAR	&	PARQP	\\ \hline \hline
100	&	5	&	ASE	&	0,0140	&	0,0169	&	0,0205	&	0,0037	&	0,0039	&	0,0054	&	0,0037	&	0,0037	&	0,0037	&	0,0161	&	0,0037	\\
	&		&	Bias	&	0,0988	&	0,1114	&	0,1249	&	0,0044	&	0,0075	&	0,0252	&	0,0033	&	0,0033	&	0,0033	&	0,1100	&	0,0033	\\
	&		&	sd	&	0,0651	&	0,0672	&	0,0699	&	0,0610	&	0,0622	&	0,0688	&	0,0604	&	0,0604	&	0,0604	&	0,0634	&	0,0604	\\ \hline
100	&	10	&		&	0,0159	&	0,0189	&	0,0225	&	0,0049	&	0,0053	&	0,0057	&	0,0039	&	0,0039	&	0,0039	&	0,0156	&	0,0039	\\
	&		&		&	0,1003	&	0,1108	&	0,1229	&	-0,0023	&	0,0013	&	0,0138	&	-0,0026	&	-0,0026	&	-0,0026	&	0,1040	&	-0,0026	\\
	&		&		&	0,0765	&	0,0812	&	0,0860	&	0,0700	&	0,0728	&	0,0744	&	0,0620	&	0,0620	&	0,0620	&	0,0692	&	0,0620	\\ \hline
250	&	5	&		&	0,0820	&	0,0858	&	0,0940	&	0,0139	&	0,0142	&	0,0147	&	0,0029	&	0,0029	&	0,0029	&	0,0570	&	0,0029	\\
	&		&		&	0,2414	&	0,2500	&	0,2646	&	0,0003	&	0,0026	&	0,0192	&	-0,0007	&	-0,0007	&	-0,0007	&	0,2245	&	-0,0007	\\
	&		&		&	0,1540	&	0,1527	&	0,1550	&	0,1181	&	0,1193	&	0,1197	&	0,0534	&	0,0534	&	0,0534	&	0,0815	&	0,0534	\\ \hline
250	&	10	&		&	0,0221	&	0,0226	&	0,0255	&	0,0021	&	0,0022	&	0,0029	&	0,0013	&	0,0013	&	0,0013	&	0,0140	&	0,0013	\\
	&		&		&	0,1246	&	0,1264	&	0,1361	&	0,0110	&	0,0130	&	0,0221	&	-0,0049	&	-0,0049	&	-0,0049	&	0,1072	&	-0,0049	\\
	&		&		&	0,0812	&	0,0813	&	0,0835	&	0,0444	&	0,0456	&	0,0489	&	0,0356	&	0,0356	&	0,0356	&	0,0496	&	0,0356	\\ \hline
500	&	5	&		&	0,1176	&	0,1196	&	0,1245	&	0,0022	&	0,0023	&	0,0025	&	0,0015	&	0,0015	&	0,0015	&	0,0578	&	0,0015	\\
	&		&		&	0,2811	&	0,2850	&	0,2939	&	0,0012	&	0,0017	&	0,0064	&	0,0011	&	0,0011	&	0,0011	&	0,2280	&	0,0011	\\
	&		&		&	0,1963	&	0,1959	&	0,1952	&	0,0474	&	0,0478	&	0,0498	&	0,0392	&	0,0392	&	0,0392	&	0,0761	&	0,0392	\\ \hline
500	&	10	&		&	0,0384	&	0,0385	&	0,0397	&	0,0011	&	0,0011	&	0,0012	&	0,0006	&	0,0006	&	0,0006	&	0,0130	&	0,0006	\\
	&		&		&	0,1462	&	0,1468	&	0,1502	&	0,0150	&	0,0155	&	0,0180	&	-0,0001	&	-0,0001	&	-0,0001	&	0,1066	&	-0,0001	\\
	&		&		&	0,1305	&	0,1303	&	0,1309	&	0,0296	&	0,0293	&	0,0304	&	0,0243	&	0,0243	&	0,0243	&	0,0400	&	0,0243	\\ \hline \hline
\end{tabular}
\caption{Simulation study, scenario 2, Gaussian responses and exponential covariate-dependent random effects. ASE, bias and standard deviation of parameter estimates}\label{table4}
\end{sidewaystable}

\begin{sidewaystable}[!ht]
\begin{tabular}{ccc|rrrrrrrrrrr}
n	&	T	&		&	FMLik	&	FMAIC	&	FMBIC	&	CovLik	&	CovAIC	&	CovBIC	&	FMQPLik	&	FMQPAIC	&	FMQPBIC	&	PAR	&	PARQP	\\ \hline \hline
100	&	5	&	ASE	&	0,0082	&	0,0093	&	0,0121	&	0,0044	&	0,0038	&	0,0035	&	0,0044	&	0,0044	&	0,0044	&	0,0063	&	0,0044	\\
	&		&	Bias	&	0,0566	&	0,0632	&	0,0779	&	0,0009	&	0,0029	&	0,0035	&	0,0024	&	0,0024	&	0,0024	&	-0,0468	&	0,0024	\\
	&		&	sd	&	0,0708	&	0,0728	&	0,0776	&	0,0664	&	0,0616	&	0,0591	&	0,0660	&	0,0660	&	0,0660	&	0,0639	&	0,0660	\\ \hline
100	&	10	&		&	0,0115	&	0,0117	&	0,0133	&	0,0035	&	0,0023	&	0,0018	&	0,0034	&	0,0034	&	0,0034	&	0,0043	&	0,0034	\\
	&		&		&	0,0857	&	0,0851	&	0,0900	&	0,0031	&	0,0032	&	0,0023	&	0,0036	&	0,0036	&	0,0036	&	-0,0319	&	0,0036	\\
	&		&		&	0,0641	&	0,0667	&	0,0718	&	0,0595	&	0,0480	&	0,0420	&	0,0584	&	0,0584	&	0,0584	&	0,0573	&	0,0584	\\ \hline
250	&	5	&		&	0,0204	&	0,0208	&	0,0253	&	0,0030	&	0,0021	&	0,0013	&	0,0029	&	0,0029	&	0,0029	&	0,0069	&	0,0029	\\
	&		&		&	0,1320	&	0,1336	&	0,1469	&	-0,0072	&	-0,0022	&	-0,0026	&	-0,0046	&	-0,0046	&	-0,0046	&	-0,0679	&	-0,0046	\\
	&		&		&	0,0542	&	0,0545	&	0,0606	&	0,0539	&	0,0460	&	0,0363	&	0,0540	&	0,0540	&	0,0540	&	0,0483	&	0,0540	\\ \hline
250	&	10	&		&	0,0097	&	0,0092	&	0,0092	&	0,0013	&	0,0010	&	0,0007	&	0,0014	&	0,0014	&	0,0014	&	0,0023	&	0,0014	\\
	&		&		&	0,0841	&	0,0820	&	0,0818	&	0,0020	&	0,0024	&	0,0016	&	0,0038	&	0,0038	&	0,0038	&	-0,0335	&	0,0038	\\
	&		&		&	0,0509	&	0,0500	&	0,0504	&	0,0366	&	0,0308	&	0,0268	&	0,0369	&	0,0369	&	0,0369	&	0,0337	&	0,0369	\\ \hline
500	&	5	&		&	0,0218	&	0,0217	&	0,0236	&	0,0015	&	0,0009	&	0,0007	&	0,0015	&	0,0015	&	0,0015	&	0,0051	&	0,0015	\\
	&		&		&	0,1407	&	0,1405	&	0,1457	&	-0,0001	&	0,0001	&	0,0011	&	-0,0001	&	-0,0001	&	-0,0001	&	-0,0626	&	-0,0001	\\
	&		&		&	0,0444	&	0,0444	&	0,0482	&	0,0384	&	0,0293	&	0,0273	&	0,0390	&	0,0390	&	0,0390	&	0,0338	&	0,0390	\\ \hline
500	&	10	&		&	0,0084	&	0,0081	&	0,0080	&	0,0006	&	0,0004	&	0,0003	&	0,0006	&	0,0006	&	0,0006	&	0,0019	&	0,0006	\\
	&		&		&	0,0825	&	0,0810	&	0,0806	&	-0,0002	&	0,0008	&	0,0012	&	0,0004	&	0,0004	&	0,0004	&	-0,0369	&	0,0004	\\
	&		&		&	0,0396	&	0,0389	&	0,0388	&	0,0248	&	0,0203	&	0,0176	&	0,0250	&	0,0250	&	0,0250	&	0,0234	&	0,0250	\\ \hline \hline
\end{tabular}
\caption{Simulation study, scenario 3, Gaussian responses and discrete random effect with covariate-dependent masses. ASE, bias and standard deviation of parameter estimates}\label{table5}
\end{sidewaystable}

Let us turn to the scenarios regarding a (conditional) Bernoulli response.
\begin{center}
\textbf{Tables \ref{table6}--\ref{table11} about here}
\end{center}

As it can be evinced by looking at Tables \ref{table6}--\ref{table8}, the \emph{standard} (misspecified) random effect estimator leads also in this case to substantially biased estimates, and the bias is increasing with correlation between the omitted (represented by the random effects) and the observed covariates. The bias is specifically high for \emph{short} individual sequences (i.e. $T=5$) and any value of the sample size $n$. When we look at our proposal, the resulting estimates usually outperform those obtained by all the other methods, especially when $T$ is small. This is true whether we consider as competitors the finite mixture-based or the parametric QP decomposition, the fixed effect (more substantially) and the bias-corrected fixed effect (at least slightly) estimators. In this case, the particularly good behaviour we have observed for the Gaussian scenarios when the number of components of the finite mixture is chosen by looking at the maximized likelihood values only (with a minimum threshold) is less evident and, in some limited cases in Table \ref{table9}, the empirical evidence is that both AIC and BIC tend to give results that are comparable to (or slightly better than) those obtained by using the maximized log-likelihood value for choosing the number of components. However, we may not observe any substantial bias or increase in ASE when using the last criterion and, therefore, we suggest to look at this criterion also in more general settings.  We must notice that, also in this setting, for scenarios reported in Table \ref{table6}--\ref{table8} ParQP results from the \emph{true} data generating model while all the remaining estimators refer to somewhat misspecified data generating processes.

\begin{sidewaystable}[!ht]
\begin{tabular}{ccc|rrrrrrrrrrrrr}
n	&	T	&		&	FMLik	&	FMAIC	&	FMBIC	&	CovLik	&	CovAIC	&	CovBIC	&	FMQPLik	&	FMQPAIC	&	FMQPBIC	&	Par	&	ParQP &	FE	&	FEbc	\\ \hline \hline
100	&	5	&	ASE	&	0,0068	&	0,0068	&	0,0068	&	0,0070	&	0,0079	&	0,0073	&	0,0077	&	0,0080	&	0,0080	&	0,0095	&	0,0083	&	0,0379	&	0,0103	\\
	&		&	Bias	&	-0,0050	&	-0,0102	&	-0,0131	&	-0,0084	&	-0,0421	&	-0,0426	&	-0,0181	&	-0,0238	&	-0,0238	&	0,0338	&	-0,0083	&	0,1615	&	0,0492	\\
	&		&	sd	&	0,0826	&	0,0818	&	0,0813	&	0,0835	&	0,0780	&	0,0740	&	0,0859	&	0,0860	&	0,0860	&	0,0914	&	0,0908	&	0,1086	&	0,0887	\\ \hline
100	&	10	&		&	0,0030	&	0,0030	&	0,0031	&	0,0033	&	0,0038	&	0,0038	&	0,0032	&	0,0031	&	0,0031	&	0,0043	&	0,0034	&	0,0076	&	0,0032	\\
	&		&		&	-0,0023	&	-0,0054	&	-0,0078	&	-0,0100	&	-0,0337	&	-0,0342	&	-0,0013	&	-0,0071	&	-0,0071	&	0,0288	&	0,0045	&	0,0619	&	0,0123	\\
	&		&		&	0,0544	&	0,0546	&	0,0551	&	0,0563	&	0,0519	&	0,0515	&	0,0564	&	0,0553	&	0,0553	&	0,0588	&	0,0584	&	0,061	&	0,0556	\\ \hline
250	&	5	&		&	0,0027	&	0,0027	&	0,0027	&	0,0023	&	0,0025	&	0,0025	&	0,0027	&	0,0028	&	0,0028	&	0,0051	&	0,0031	&	0,0293	&	0,0057	\\
	&		&		&	0,0007	&	-0,0026	&	-0,0039	&	-0,0039	&	-0,0140	&	-0,0170	&	-0,0088	&	-0,0127	&	-0,0127	&	0,0411	&	0,0025	&	0,153	&	0,0425	\\
	&		&		&	0,0515	&	0,0522	&	0,0521	&	0,0473	&	0,0475	&	0,0471	&	0,0516	&	0,0512	&	0,0512	&	0,0583	&	0,0557	&	0,0771	&	0,0627	\\ \hline
250	&	10	&		&	0,0011	&	0,0012	&	0,0012	&	0,0011	&	0,0013	&	0,0013	&	0,0012	&	0,0012	&	0,0012	&	0,0020	&	0,0012	&	0,0055	&	0,0015	\\
	&		&		&	-0,0054	&	-0,0070	&	-0,0082	&	-0,0032	&	-0,0167	&	-0,0171	&	-0,0049	&	-0,0074	&	-0,0074	&	0,0266	&	0,0017	&	0,0615	&	0,0119	\\
	&		&		&	0,0329	&	0,0332	&	0,0333	&	0,0331	&	0,0314	&	0,0311	&	0,0338	&	0,0339	&	0,0339	&	0,0366	&	0,0351	&	0,0409	&	0,0372	\\ \hline
500	&	5	&		&	0,0015	&	0,0016	&	0,0016	&	0,0012	&	0,0016	&	0,0016	&	0,0014	&	0,0015	&	0,0015	&	0,0033	&	0,0014	&	0,0247	&	0,0032	\\
	&		&		&	-0,0043	&	-0,0069	&	-0,0086	&	-0,0123	&	-0,0253	&	-0,0259	&	-0,0135	&	-0,0167	&	-0,0167	&	0,0364	&	-0,0018	&	0,149	&	0,0395	\\
	&		&		&	0,0389	&	0,0393	&	0,0394	&	0,0328	&	0,0315	&	0,0311	&	0,0356	&	0,0353	&	0,0353	&	0,0447	&	0,0379	&	0,0497	&	0,0407	\\ \hline
500	&	10	&		&	0,0006	&	0,0006	&	0,0007	&	0,0004	&	0,0004	&	0,0004	&	0,0005	&	0,0006	&	0,0006	&	0,0013	&	0,0005	&	0,0247	&	0,0032	\\
	&		&		&	-0,0084	&	-0,0091	&	-0,0103	&	-0,0011	&	-0,0046	&	-0,0048	&	-0,0063	&	-0,0077	&	-0,0077	&	0,0233	&	0,0007	&	0,149	&	0,0395	\\
	&		&		&	0,0233	&	0,0234	&	0,0238	&	0,0210	&	0,0207	&	0,0204	&	0,0224	&	0,0223	&	0,0223	&	0,0275	&	0,0233	&	0,0497	&	0,0407	\\ \hline \hline
\end{tabular}
\caption{Simulation study, scenario 1.1, Bernoulli responses and Gaussian random effects, constant correlation $\rho={\rm cor}\left(u_{i}, x_{it}\right) \in \left(0, 0.2 \right]$. ASE, bias and standard deviation of parameter estimates}\label{table6}
\end{sidewaystable}

\begin{sidewaystable}[!ht]
\begin{tabular}{ccc|rrrrrrrrrrrrr}
n	&	T	&		&	FMLik	&	FMAIC	&	FMBIC	&	CovLik	&	CovAIC	&	CovBIC	&	FMQPLik	&	FMQPAIC	&	FMQPBIC	&	Par	&	ParQP &	FE	&	FEbc	\\ \hline \hline
100	&	5	&	ASE	&	0,0185	&	0,0178	&	0,0185	&	0,0072	&	0,0063	&	0,0062	&	0,0079	&	0,0080	&	0,0080	&	0,0357	&	0,0089	&	0,0381	&	0,0125	\\
	&		&	Bias	&	0,1030	&	0,0996	&	0,1012	&	-0,0253	&	-0,0163	&	-0,0165	&	0,0033	&	-0,0039	&	-0,0039	&	0,1605	&	0,0119	&	0,1469	&	0,0371	\\
	&		&	sd	&	0,0890	&	0,0888	&	0,0909	&	0,0807	&	0,0774	&	0,0773	&	0,0888	&	0,0893	&	0,0893	&	0,0998	&	0,0935	&	0,1284	&	0,1057	\\ \hline
100	&	10	&		&	0,0076	&	0,0078	&	0,0080	&	0,0046	&	0,0043	&	0,0043	&	0,0051	&	0,0051	&	0,0051	&	0,0187	&	0,0053	&	0,0093	&	0,0043	\\
	&		&		&	0,0567	&	0,0578	&	0,0594	&	-0,0250	&	-0,0182	&	-0,0184	&	-0,0056	&	-0,0108	&	-0,0108	&	0,1122	&	-0,0020	&	0,0669	&	0,0165	\\
	&		&		&	0,0664	&	0,0670	&	0,0668	&	0,0629	&	0,0629	&	0,0628	&	0,0712	&	0,0706	&	0,0706	&	0,0780	&	0,0728	&	0,0697	&	0,0634	\\ \hline
250	&	5	&		&	0,0157	&	0,0152	&	0,0151	&	0,0033	&	0,0031	&	0,0031	&	0,0033	&	0,0033	&	0,0033	&	0,0333	&	0,0037	&	0,0265	&	0,0058	\\
	&		&		&	0,1047	&	0,1023	&	0,1009	&	-0,0317	&	-0,0215	&	-0,0218	&	-0,0007	&	-0,0045	&	-0,0045	&	0,1652	&	0,0087	&	0,1396	&	0,0325	\\
	&		&		&	0,0689	&	0,0691	&	0,0699	&	0,0481	&	0,0515	&	0,0513	&	0,0573	&	0,0577	&	0,0577	&	0,0777	&	0,0602	&	0,0836	&	0,0692	\\ \hline
250	&	10	&		&	0,0049	&	0,0050	&	0,0054	&	0,0014	&	0,0015	&	0,0014	&	0,0017	&	0,0017	&	0,0017	&	0,0150	&	0,0017	&	0,0057	&	0,0019	\\
	&		&		&	0,0520	&	0,0529	&	0,0549	&	-0,0087	&	-0,0033	&	-0,0035	&	-0,0098	&	-0,0125	&	-0,0125	&	0,1084	&	-0,0053	&	0,0598	&	0,0104	\\
	&		&		&	0,0468	&	0,0471	&	0,0486	&	0,0368	&	0,0389	&	0,0370	&	0,0396	&	0,0394	&	0,0394	&	0,0569	&	0,0405	&	0,0464	&	0,0423	\\ \hline
500	&	5	&		&	0,0326	&	0,0322	&	0,0318	&	0,0181	&	0,0175	&	0,0174	&	0,0220	&	0,0219	&	0,0219	&	0,0537	&	0,0245	&	0,0265	&	0,0044	\\
	&		&		&	0,1062	&	0,1044	&	0,1034	&	-0,0369	&	-0,0203	&	-0,0205	&	-0,0048	&	-0,0078	&	-0,0078	&	0,1671	&	0,0068	&	0,1506	&	0,0421	\\
	&		&		&	0,1460	&	0,1458	&	0,1454	&	0,1294	&	0,1309	&	0,1303	&	0,1483	&	0,1479	&	0,1479	&	0,1605	&	0,1563	&	0,0614	&	0,0512	\\ \hline
500	&	10	&		&	0,0045	&	0,0045	&	0,0047	&	0,0019	&	0,0016	&	0,0017	&	0,0009	&	0,0009	&	0,0009	&	0,0150	&	0,0010	&	0,0056	&	0,0012	\\
	&		&		&	0,0560	&	0,0560	&	0,0574	&	-0,0355	&	-0,0308	&	-0,0311	&	-0,0053	&	-0,0069	&	-0,0069	&	0,1130	&	-0,0003	&	0,0668	&	0,0167	\\
	&		&		&	0,0374	&	0,0375	&	0,0381	&	0,0253	&	0,0264	&	0,0262	&	0,0302	&	0,0300	&	0,0300	&	0,0470	&	0,0310	&	0,0341	&	0,0311	\\ \hline \hline
\end{tabular}
\caption{Simulation study, scenario 1.2, Bernoulli responses and Gaussian random effects, constant correlation $\rho={\rm cor}\left(u_{i}, x_{it}\right) \in \left(0.2, 0.5 \right]$. ASE, bias and standard deviation of parameter estimates}\label{table7}
\end{sidewaystable}

\begin{sidewaystable}[!ht]
\begin{tabular}{ccc|rrrrrrrrrrrrr}
n	&	T	&		&	FMLik	&	FMAIC	&	FMBIC	&	CovLik	&	CovAIC	&	CovBIC	&	FMQPLik	&	FMQPAIC	&	FMQPBIC	&	Par	&	ParQP &	FE	&	FEbc	\\ \hline \hline
100	&	5	&	ASE	&	0,1258	&	0,1329	&	0,1428	&	0,0238	&	0,0312	&	0,0311	&	0,0357	&	0,0339	&	0,0339	&	0,1804	&	0,0375	&	0,0563	&	0,0237	\\
	&		&	Bias	&	0,3073	&	0,3152	&	0,3307	&	0,0063	&	0,0859	&	0,0861	&	-0,0041	&	-0,0177	&	-0,0177	&	0,3838	&	0,0009	&	0,1588	&	0,0468	\\
	&		&	sd	&	0,1770	&	0,1833	&	0,1829	&	0,1542	&	0,1543	&	0,1539	&	0,1888	&	0,1832	&	0,1832	&	0,1820	&	0,1937	&	0,1762	&	0,1466	\\ \hline
250	&	10	&		&	0,0547	&	0,0564	&	0,0579	&	0,0022	&	0,0051	&	0,0051	&	0,0034	&	0,0034	&	0,0034	&	0,1181	&	0,0035	&	0,0149	&	0,009	\\
	&		&		&	0,2197	&	0,2231	&	0,2265	&	0,0054	&	0,0437	&	0,0441	&	0,0037	&	0,0013	&	0,0013	&	0,3295	&	0,0045	&	0,0659	&	0,0152	\\
	&		&		&	0,0799	&	0,0811	&	0,0815	&	0,0470	&	0,0562	&	0,0558	&	0,0583	&	0,0582	&	0,0582	&	0,0978	&	0,0591	&	0,103	&	0,0936	\\ \hline
250	&	5	&		&	0,0857	&	0,0855	&	0,0905	&	0,0031	&	0,0070	&	0,0070	&	0,0062	&	0,0063	&	0,0063	&	0,1372	&	0,0064	&	0,0354	&	0,011	\\
	&		&		&	0,2794	&	0,2786	&	0,2844	&	-0,0075	&	0,0518	&	0,0521	&	-0,0113	&	-0,0157	&	-0,0157	&	0,3590	&	-0,0055	&	0,1468	&	0,0377	\\
	&		&		&	0,0874	&	0,0889	&	0,0980	&	0,0553	&	0,0658	&	0,0654	&	0,0778	&	0,0775	&	0,0775	&	0,0911	&	0,0797	&	0,1177	&	0,098	\\ \hline
250	&	10	&		&	0,0504	&	0,0518	&	0,0534	&	0,0019	&	0,0056	&	0,0055	&	0,0033	&	0,0032	&	0,0032	&	0,1117	&	0,0034	&	0,008	&	0,0038	\\
	&		&		&	0,2127	&	0,2155	&	0,2191	&	0,0044	&	0,0454	&	0,0458	&	0,0081	&	0,0055	&	0,0055	&	0,3217	&	0,0091	&	0,0596	&	0,0098	\\
	&		&		&	0,0720	&	0,0730	&	0,0734	&	0,0429	&	0,0591	&	0,0585	&	0,0565	&	0,0563	&	0,0563	&	0,0906	&	0,0572	&	0,0667	&	0,0607	\\ \hline
500	&	5	&		&	0,0943	&	0,0945	&	0,0960	&	0,0021	&	0,0052	&	0,0052	&	0,0033	&	0,0033	&	0,0033	&	0,1473	&	0,0034	&	0,0287	&	0,0059	\\
	&		&		&	0,2940	&	0,2938	&	0,2956	&	-0,0009	&	0,0436	&	0,0440	&	-0,0086	&	-0,0108	&	-0,0108	&	0,3735	&	-0,0014	&	0,1505	&	0,0412	\\
	&		&		&	0,0887	&	0,0903	&	0,0929	&	0,0463	&	0,0574	&	0,0572	&	0,0565	&	0,0561	&	0,0561	&	0,0885	&	0,0579	&	0,0774	&	0,0648	\\ \hline
500	&	10	&		&	0,0475	&	0,0482	&	0,0495	&	0,0010	&	0,0018	&	0,0018	&	0,0015	&	0,0015	&	0,0015	&	0,1075	&	0,0015	&	0,0062	&	0,002	\\
	&		&		&	0,2070	&	0,2085	&	0,2115	&	-0,0019	&	0,0191	&	0,0193	&	-0,0048	&	-0,0063	&	-0,0063	&	0,3161	&	-0,0034	&	0,0639	&	0,0138	\\
	&		&		&	0,0683	&	0,0684	&	0,0692	&	0,0308	&	0,0383	&	0,0380	&	0,0384	&	0,0383	&	0,0383	&	0,0873	&	0,0392	&	0,0465	&	0,0423	\\ \hline \hline
\end{tabular}
\caption{Simulation study, scenario 1.3, Bernoulli responses and Gaussian random effects, constant correlation $\rho={\rm cor}\left(u_{i}, x_{it}\right) \in \left(0.5, 0.8 \right]$. ASE, bias and standard deviation of parameter estimates}\label{table8}
\end{sidewaystable}

\begin{sidewaystable}[!ht]
\begin{tabular}{ccc|rrrrrrrrrrrrr}
n	&	T	&		&	FMLik	&	FMAIC	&	FMBIC	&	CovLik	&	CovAIC	&	CovBIC	&	FMQPLik	&	FMQPAIC	&	FMQPBIC	&	Par	&	ParQP &	FE	&	FEbc	\\ \hline \hline
100	&	5	&	ASE	&	0,0473	&	0,0425	&	0,0413	&	0,0167	&	0,0096	&	0,0096	&	0,0238	&	0,0230	&	0,0230	&	0,1595	&	0,0264	&	0,0541	&	0,0189	\\
	&		&	Bias	&	0,1863	&	0,1744	&	0,1714	&	-0,0112	&	0,0174	&	0,0174	&	-0,0143	&	-0,0271	&	-0,0271	&	0,3685	&	0,0016	&	0,1765	&	0,0603	\\
	&		&	sd	&	0,1120	&	0,1101	&	0,1091	&	0,1287	&	0,0964	&	0,0964	&	0,1535	&	0,1491	&	0,1491	&	0,1541	&	0,1624	&	0,1514	&	0,1236	\\ \hline
100	&	10	&		&	0,0134	&	0,0137	&	0,0145	&	0,0070	&	0,0053	&	0,0053	&	0,0091	&	0,0091	&	0,0091	&	0,0704	&	0,0096	&	0,0126	&	0,0063	\\
	&		&		&	0,0880	&	0,0881	&	0,0915	&	-0,0151	&	-0,0115	&	-0,0113	&	-0,0068	&	-0,0144	&	-0,0144	&	0,2430	&	0,0022	&	0,0743	&	0,0225	\\
	&		&		&	0,0752	&	0,0768	&	0,0784	&	0,0822	&	0,0718	&	0,0716	&	0,0954	&	0,0943	&	0,0943	&	0,1066	&	0,0982	&	0,0841	&	0,0763	\\ \hline
250	&	5	&		&	0,0398	&	0,0372	&	0,0362	&	0,0057	&	0,0042	&	0,0041	&	0,0095	&	0,0096	&	0,0096	&	0,1534	&	0,0103	&	0,0337	&	0,008	\\
	&		&		&	0,1871	&	0,1799	&	0,1767	&	-0,0217	&	0,0004	&	0,0006	&	-0,0175	&	-0,0253	&	-0,0253	&	0,3777	&	-0,0005	&	0,1586	&	0,0468	\\
	&		&		&	0,0694	&	0,0697	&	0,0706	&	0,0720	&	0,0647	&	0,0643	&	0,0958	&	0,0947	&	0,0947	&	0,1035	&	0,1016	&	0,0925	&	0,0765	\\ \hline
250	&	10	&		&	0,0115	&	0,0113	&	0,0117	&	0,0029	&	0,0026	&	0,0026	&	0,0044	&	0,0044	&	0,0044	&	0,0727	&	0,0046	&	0,0077	&	0,0028	\\
	&		&		&	0,0927	&	0,0915	&	0,0926	&	-0,0179	&	-0,0211	&	-0,0211	&	-0,0058	&	-0,0097	&	-0,0097	&	0,2544	&	0,0044	&	0,0681	&	0,0171	\\
	&		&		&	0,0538	&	0,0540	&	0,0559	&	0,0504	&	0,0464	&	0,0464	&	0,0658	&	0,0654	&	0,0654	&	0,0895	&	0,0679	&	0,0553	&	0,0502	\\ \hline
500	&	5	&		&	0,0388	&	0,0372	&	0,0349	&	0,0021	&	0,0020	&	0,0020	&	0,0046	&	0,0047	&	0,0049	&	0,1538	&	0,0050	&	0,0284	&	0,0054	\\
	&		&		&	0,1886	&	0,1842	&	0,1781	&	-0,0056	&	-0,0012	&	-0,0012	&	-0,0144	&	-0,0189	&	-0,0252	&	0,3808	&	0,0040	&	0,1515	&	0,0412	\\
	&		&		&	0,0570	&	0,0575	&	0,0567	&	0,0450	&	0,0450	&	0,0450	&	0,0664	&	0,0662	&	0,0653	&	0,0938	&	0,0706	&	0,0737	&	0,0607	\\ \hline
500	&	10	&		&	0,0101	&	0,0100	&	0,0101	&	0,0012	&	0,0011	&	0,0012	&	0,0023	&	0,0023	&	0,0023	&	0,0689	&	0,0025	&	0,0074	&	0,002	\\
	&		&		&	0,0914	&	0,0909	&	0,0916	&	-0,0067	&	-0,0087	&	-0,0090	&	-0,0052	&	-0,0076	&	-0,0099	&	0,2485	&	0,0054	&	0,0753	&	0,0236	\\
	&		&		&	0,0421	&	0,0421	&	0,0416	&	0,0344	&	0,0327	&	0,0336	&	0,0478	&	0,0476	&	0,0468	&	0,0844	&	0,0493	&	0,0417	&	0,0378	\\ \hline \hline
\end{tabular}
\caption{Simulation study, scenario 2, Bernoulli responses and exponential covariate-dependent random effects. ASE, bias and standard deviation of parameter estimates}\label{table9}
\end{sidewaystable}

\begin{sidewaystable}[!ht]
\begin{tabular}{ccc|rrrrrrrrrrrrr}
n	&	T	&		&	FMLik	&	FMAIC	&	FMBIC	&	CovLik	&	CovAIC	&	CovBIC	&	FMQPLik	&	FMQPAIC	&	FMQPBIC	&	Par	&	ParQP &	FE	&	FEbc	\\ \hline \hline
100	&	5	&	ASE	&	0,0362	&	0,0379	&	0,0431	&	0,0198	&	0,0161	&	0,0161	&	0,0350	&	0,0322	&	0,0316	&	0,3859	&	0,0470	&	0,1843	&	0,0654	\\
	&		&	Bias	&	0,1420	&	0,1486	&	0,1626	&	0,0083	&	-0,0420	&	-0,0422	&	0,0122	&	-0,0094	&	-0,0162	&	0,5825	&	0,0607	&	0,3365	&	0,1554	\\
	&		&	sd	&	0,1266	&	0,1256	&	0,1291	&	0,1405	&	0,1199	&	0,1198	&	0,1866	&	0,1792	&	0,1770	&	0,2158	&	0,2081	&	0,2666	&	0,203	\\ \hline
100	&	10	&		&	0,0122	&	0,0118	&	0,0112	&	0,0084	&	0,0090	&	0,0089	&	0,0150	&	0,0149	&	0,0148	&	0,1229	&	0,0181	&	0,0335	&	0,0153	\\
	&		&		&	0,0644	&	0,0611	&	0,0524	&	-0,0151	&	-0,0257	&	-0,0259	&	0,0009	&	-0,0022	&	-0,0085	&	0,3190	&	0,0292	&	0,1386	&	0,0656	\\
	&		&		&	0,0897	&	0,0899	&	0,0919	&	0,0906	&	0,0911	&	0,0909	&	0,1224	&	0,1220	&	0,1215	&	0,1453	&	0,1312	&	0,1196	&	0,1051	\\ \hline
250	&	5	&		&	0,0227	&	0,0224	&	0,0266	&	0,0058	&	0,0071	&	0,0071	&	0,0123	&	0,0121	&	0,0115	&	0,3440	&	0,0181	&	0,1016	&	0,0297	\\
	&		&		&	0,1235	&	0,1222	&	0,1346	&	0,0040	&	-0,0182	&	-0,0184	&	0,0053	&	-0,0027	&	-0,0138	&	0,5616	&	0,0541	&	0,2816	&	0,129	\\
	&		&		&	0,0865	&	0,0862	&	0,0923	&	0,0761	&	0,0825	&	0,0824	&	0,1110	&	0,1098	&	0,1063	&	0,1691	&	0,1231	&	0,1495	&	0,1141	\\ \hline
250	&	10	&		&	0,0077	&	0,0074	&	0,0072	&	0,0029	&	0,0033	&	0,0029	&	0,0076	&	0,0075	&	0,0076	&	0,1143	&	0,0087	&	0,03	&	0,0112	\\
	&		&		&	0,0675	&	0,0653	&	0,0622	&	-0,0037	&	-0,0015	&	-0,0016	&	-0,0107	&	-0,0142	&	-0,0147	&	0,3224	&	0,0203	&	0,1543	&	0,0798	\\
	&		&		&	0,0559	&	0,0563	&	0,0575	&	0,0535	&	0,0570	&	0,0540	&	0,0867	&	0,0857	&	0,0858	&	0,1016	&	0,0913	&	0,0789	&	0,0694	\\ \hline
500	&	5	&		&	0,0180	&	0,0172	&	0,0186	&	0,0024	&	0,0027	&	0,0027	&	0,0077	&	0,0076	&	0,0075	&	0,3316	&	0,0133	&	0,0891	&	0,0241	\\
	&		&		&	0,1210	&	0,1178	&	0,1230	&	0,0061	&	0,0002	&	0,0003	&	0,0076	&	0,0033	&	-0,0054	&	0,5597	&	0,0580	&	0,2794	&	0,1311	\\
	&		&		&	0,0581	&	0,0578	&	0,0592	&	0,0485	&	0,0521	&	0,0518	&	0,0875	&	0,0871	&	0,0862	&	0,1354	&	0,0997	&	0,1052	&	0,0831	\\ \hline
500	&	10	&		&	0,0062	&	0,0060	&	0,0058	&	0,0011	&	0,0013	&	0,0013	&	0,0028	&	0,0027	&	0,0028	&	0,1174	&	0,0046	&	0,0218	&	0,0067	\\
	&		&		&	0,0665	&	0,0658	&	0,0641	&	-0,0049	&	0,0027	&	0,0029	&	0,0039	&	0,0030	&	0,0022	&	0,3310	&	0,0364	&	0,1364	&	0,0647	\\
	&		&		&	0,0416	&	0,0412	&	0,0413	&	0,0335	&	0,0361	&	0,0359	&	0,0525	&	0,0517	&	0,0525	&	0,0885	&	0,0569	&	0,0566	&	0,0501	\\ \hline \hline
\end{tabular}
\caption{Simulation study, scenario 3, Bernoulli responses and discrete random effect with covariate-dependent masses. ASE, bias and standard deviation of parameter estimates}\label{table10}
\end{sidewaystable}

\begin{sidewaystable}[!ht]
\begin{tabular}{ccc|rrrrrrrrrrrrr}
n	&	T	&		&	FMLik	&	FMAIC	&	FMBIC	&	CovLik	&	CovAIC	&	CovBIC	&	FMQPLik	&	FMQPAIC	&	FMQPBIC	&	Par	&	ParQP &	FE	&	FEbc	\\ \hline \hline
100	&	5	&	ASE	&	0.2736	&	0.2968	&	0.2988	&	0.1204	&	0.1229	&	0.1377	&	0.1372	&	0.1356	&	0.1358	&	0.3069	&	0.1381	&	0.2017	&	0.1538	\\
	&		&	Bias	&	0.4049	&	0.4294	&	0.4396	&	0.0153	&	0.0745	&	0.136	&	-0.012	&	-0.0216	&	-0.0222	&	0.4481	&	-0.0116	&	0.1349	&	0.0280	\\
	&		&	sd	&	0.1097	&	0.1124	&	0.1056	&	0.1202	&	0.1173	&	0.1192	&	0.1371	&	0.1351	&	0.1353	&	0.1061	&	0.1380	&	0.1835	&	0.1530	\\ \hline
100	&	10	&		&	0.1784	&	0.1876	&	0.2194	&	0.0823	&	0.0868	&	0.0979	&	0.0904	&	0.0886	&	0.0878	&	0.2587	&	0.0895	&	0.1081	&	0.0953	\\
	&		&		&	0.2977	&	0.3037	&	0.3267	&	0.0099	&	0.0509	&	0.0970	&	-0.0042	&	-0.0131	&	-0.0177	&	0.3969	&	-0.0093	&	0.0598	&	0.0101	\\
	&		&		&	0.0898	&	0.0954	&	0.1127	&	0.0822	&	0.0842	&	0.0885	&	0.0904	&	0.0884	&	0.0875	&	0.1012	&	0.0894	&	0.1045	&	0.0952	\\ \hline
250	&	5	&		&	0.2484	&	0.2659	&	0.2747	&	0.0695	&	0.0722	&	0.0932	&	0.0776	&	0.0777	&	0.0779	&	0.2829	&	0.0791	&	0.1315	&	0.0916	\\
	&		&		&	0.3984	&	0.4114	&	0.4267	&	0.0205	&	0.0565	&	0.1122	&	-0.0011	&	-0.010	&	-0.0112	&	0.4427	&	0.0014	&	0.1566	&	0.0466	\\
	&		&		&	0.0897	&	0.0967	&	0.0926	&	0.0691	&	0.0690	&	0.0806	&	0.0776	&	0.0776	&	0.0778	&	0.0869	&	0.0791	&	0.1070	&	0.0894	\\ \hline
250	&	10	&		&	0.1577	&	0.1591	&	0.1627	&	0.0479	&	0.0569	&	0.0722	&	0.0525	&	0.0532	&	0.0525	&	0.2313	&	0.0523	&	0.0640	&	0.0549	\\
	&		&		&	0.2848	&	0.2861	&	0.2882	&	0.0072	&	0.0284	&	0.0699	&	-0.0031	&	-0.0064	&	-0.0143	&	0.3797	&	-0.0071	&	0.0622	&	0.0125	\\
	&		&		&	0.0766	&	0.0772	&	0.0796	&	0.0478	&	0.0561	&	0.0673	&	0.0525	&	0.0532	&	0.0523	&	0.0871	&	0.0522	&	0.0601	&	0.0547	\\ \hline
500	&	5	&		&	0.2359	&	0.2503	&	0.2631	&	0.0477	&	0.0525	&	0.0664	&	0.0552	&	0.0552	&	0.0554	&	0.2672	&	0.0561	&	0.0945	&	0.0641	\\
	&		&		&	0.4060	&	0.4137	&	0.4288	&	0.0123	&	0.0374	&	0.0809	&	-0.0149	&	-0.0220	&	-0.0227	&	0.4492	&	-0.0122	&	0.1372	&	0.0302	\\
	&		&		&	0.0711	&	0.0792	&	0.0792	&	0.0475	&	0.0511	&	0.0599	&	0.0550	&	0.0547	&	0.0549	&	0.0654	&	0.0560	&	0.0757	&	0.0632	\\ \hline
500	&	10	&		&	0.1498	&	0.1513	&	0.1551	&	0.0334	&	0.0376	&	0.0429	&	0.0384	&	0.0384	&	0.0389	&	0.2246	&	0.0385	&	0.0483	&	0.0409	\\
	&		&		&	0.2909	&	0.2931	&	0.2953	&	0.0058	&	0.0194	&	0.0483	&	-0.0053	&	-0.0069	&	-0.0116	&	0.3868	&	-0.0093	&	0.0597	&	0.0103	\\
	&		&		&	0.0652	&	0.0654	&	0.0679	&	0.0334	&	0.0372	&	0.0406	&	0.0384	&	0.0384	&	0.0388	&	0.0750	&	0.0384	&	0.0447	&	0.0408	\\ \hline \hline
\end{tabular}
\caption{Simulation study, scenario 4, Bernoulli responses and continuous random effects as function of $\max_t{\bf x}_{it}$. ASE, bias and standard deviation of parameter estimates}\label{table11}
\end{sidewaystable}
As for table \ref{table11}, we may observe that the proposed estimator outperforms (sometimes slightly) all the others, especially when the number of components is selected by looking at the maximized likelihood values.

\section{A benchmark data example}\label{empdata}
In this section, we propose the re-analysis of the well known \emph{Union} data, see \cite{stat:2017}. The data come from a longitudinal, observational, study, the US National Longitudinal Survey of Mature and Young Women (NLSW, see {\url https://www.nlsinfo.org/content/cohorts/mature-and-young-women}). According to the sample definition available on the US Bureau of Labour website, the NLSW is a two-cohort survey, defined as a part of the NLS Original Cohort project, including 5,083 women who were aged 30-44 when first interviewed in 1967, and 5,159 women who were aged 14-24 when first interviewed in 1968. Data are available through 2003, when the survey was discontinued. The dataset we are currently analyzing is composed by 4,434 employed women aged 14--24 in 1968. Data refer to interviews taken between 1970 and 1988; due to the length of the study, most of the participants to the initial wave have not been interviewed in each of the measurement occasions, thus leading to an unbalanced study. The \emph{global} sample size is 26,200, with a non constant number of measurements for each unit, $T_{i} \in [1,12]$; the median number of measurements is equal to 6, while the corresponding mean is 5.94. A huge literature has focused on the determinants of union membership. Just referring back to the 80s, we may cite the seminal papers by \cite{farb:1983} and \cite{leig:1985}. With time passing by, the current dataset has been used as a benchmark to describe regression models for binary longitudinal responses. 
The response variable is \emph{union} (union membership) and, for each subject and each time occasion, the following covariates are available
\begin{itemize}
\item Year: interview year, ranging in (19)70 -- (19)88
\item Age: age in current year, ranging in 16 -- 46,
\item Grade: years of schooling completed, ranging from 0 to 18),
\item NS: living  outside  a  standard  metropolitan  statistical  area, SMSA
\item South: living in the South
\item Black: afro-american
\end{itemize}

A random intercept model is specified as follows:
\begin{eqnarray*}
g\left[{\rm E}(union_{it} \mid {\bf x}_{i}, u_{i})\right] & = & u_{i}+\beta_{1} year_{t} + \beta_{2} age_{it} + \beta_{3} grade_{it} + \\
& + & \beta_{4} ns_{it} + \beta_{5} south_{it} + \beta_{6} south_{it}*year_{t}
\end{eqnarray*}
where $u_{i}$ represents an individual-specific, time-constant, intercept, with (unknown) density $f_{U}$. We use a logit link, that is, $g(x)=\log(x/(1-x))$, and compare different estimators of $\bbeta$:
\begin{itemize}
\item the proposed finite mixture model with covariates-dependent priors (COV) by choosing the number of components either by AIC, \cite{akai:1973}, or BIC, \cite{schw:1978};
\item a standard parametric mixed model without (Par) or with QP decomposition (ParQP), fitted by using adaptive quadrature schemes, see \cite{pine:bate:1995}, and 
\item a bias-corrected fixed effect estimator (FEbc), available in the R package \texttt{bife}, see \cite{stam:2016}.
\end{itemize} 

The aim of the analysis is to explore potential determinants of union membership for young women.  As discussed by \cite{leig:1985}, some empirical evidence seems to suggest that the probability of union membership depends on, among other factors, race, region and occupation. The topic has attracted a huge literature, as it can be evinced by looking at one of the latest contributions to the field, see \cite{fran:2015}. In Table \ref{risultati}, we report the estimates for the \emph{within} and \emph{between} effects (for the QPPar specification). The former refer to the estimates associated to the dynamics in the observed covariates, while the latter refer to the estimates for the average covariates' values in the QPPar specification.

\begin{table}
\setlength\tabcolsep{2pt}
\hskip-1cm\begin{tabular}{l| r l | r l | r l | r l | r l |}
\hline
&	\multicolumn{2}{c|}{FEbc}			&	\multicolumn{2}{c|}{Par}			&	\multicolumn{2}{c|}{QPPar}			&	\multicolumn{2}{c|}{CovBIC($K=4$)}			&	\multicolumn{2}{c|}{CovAIC($K=6$)} \\ \hline	
{\bf Within}	&	beta	&	se	&	beta	&	se	&	beta	&	se	&	beta	&	se	&	beta	&	se \\ \hline
Year	&	-0.064	&	0.105	&	-0.004	&	0.016	&	-0.053	&	0.096	&	-0.008	&	0.021	&	-0.044	&	0.036\\
Age	&	0.072	&	0.104	&	0.013	&	0.015	&	0.062	&	0.096	&	0.016	&	0.021	&	0.053	&	0.035\\
Grade	&	0.082 (*)	&	0.046	&	0.086 (**)	&	0.018	&	0.082 (*)	&	0.044	&	0.052 (*)	&	0.027	&	0.018	&	0.015\\	
NS	&	0.023	&	0.123	&	-0.262 (**)	&	0.083	&	0.006	&	0.116	&	-0.018	&	0.098	&	-0.103	&	0.106 \\	
South	&	-2.870 (**)	&	0.724	&	-2.775 (**)	&	0.640	&	-2.628 (**)	&	0.680	&	-2.676 (**)	&	0.661	&	-2.711 (**)	&  0.671 \\	 
South*Year	&	0.027 (**)	&	0.009	&	0.023 (**)	&	0.008	&	0.023 (**)	&	0.008	&	0.026 (**)	&	0.008	&	0.027 (**)	& 0.008 \\ \hline
{\bf Between}	&		&		&		&		&		&		&		&		&		& \\
Year	&		&		&		&		&	0.003	&	0.021	&		&		&		& \\
Age	&		&		&		&		&	0.012	&	0.015	&		&		&		& \\
Grade	&		&		&		&		&	0.078 (**)	&	0.020	&		&		&		& \\	
NS	&		&		&		&		&	-0.510 (**)	&	0.118	&		&		&		& \\	
South	&		&		&		&		&	-1.561	&	2.057	&		&		&		& \\	
South*Year	&		&		&		&		&	0.007	&	0.026	&		&		&		& \\ \hline
\end{tabular}
\caption{Union data. Longitudinal model, parameter estimates. (*) and (**) denote \emph{weakly} ($0.1\leq p \leq 0.05$) and \emph{strongly} ($p \leq 0.05$) significant effects, respectively.} \label{risultati}
\end{table}

The standard errors for the parameter estimates in the proposed model specification have been calculated by using the \cite{oake:1999} formula and the sandwich estimator, with the aim to produce robust confidence intervals, see eg \cite{roya:1986}

As we may observe, the estimates for the marginal effect of $South$ and for the interaction $South*Year$ are been found to be \emph{strongly} statistical significant for all model specification, suggesting a clear geographical gradient which is, however, decreasing with time passing by. In the $Par$, also the estimates for $Grade$ and $NS$ have been found to be \emph{strongly} significant. This result may be due to this specification not considering the difference with the corresponding \emph{between} estimates that are found to be strongly significant when considering the potential dependence of individual random intercepts on the observed covariates, as in $ParQP$. The $FEbc$, $ParQP$, $CovBIC$, $CovAIC$ usually agree on significance of parameter estimates, but for $Grade$ that is found to be \emph{weakly} significant by all specifications but the proposed model with the number of components chosen throuogh AIC, \cite{akai:1973}.

When we consider between estimates, only those for $Grade$ and $NS$ declared to be \emph{strongly} significant in $ParQP$; we  may observe that, for the proposed model specification, a sort of \emph{between} effects may be derived by looking at estimates of the impact that observed covariates mean have on the prior probabilities of component membership. In Tables \ref{RIS_pi_AIC}-\ref{RIS_pi_BIC}, we report such estimates based on either using the BIC \cite{schw:1978} or the AIC \cite{akai:1973} to choose the number of components. As we may observe, we have component-specific estimates associated to covariates individual means.

\begin{table}
\setlength\tabcolsep{4pt}
\begin{center}
\begin{tabular}{l| r r r r r r|}
\hline
{\bf Variable}	&	Comp.1&	Comp.2 &	Comp.3  & Comp.4 &	Comp.5 & Comp.6 (ref) \\ \hline
Location($\widehat{\zeta}_{k})$ & -2.447 & -1.870 & 0.503 & 2.126 & 4.128 & 4.254\\ \hline
Intercept & 10.595 & 0.593 & 3.098 & 2.823 & -3.083 & \\
Year & -0.163 (**) & -0.063 & -0.041 & -0.088 & -0.177 (**) & \\
Age & -0.024 & 0.137 (**) & 0.015 & 0.059 & 0.135 & \\
Grade & 0.325 (**) & 0.139 (**) & 0.105 (*) & 0.266 (**) & 0.905 (**) & \\
NS & -0.181 & 0.949 (**) & 0.263 & 0.251 & 1.158 & \\
South & 2.011 & -0.025 & -2.510 & -2.984 & 0.907 & \\
South*Year & -0.036	& 0.009 & 0.038 & 0.039 & -2.297 & \\ \hline
\end{tabular}
\end{center}
\caption{Union data. Prior Probability Model, parameter estimates. AIC-based solution, $K=6$ components. $(*)$ and $(**)$ denote \emph{weakly} ($0.1 \leq p \leq 0.05$) and \emph{strongly} ($p\leq0.05$) significant effects, respectively.} \label{RIS_pi_AIC}
\end{table}

\begin{table}
\setlength\tabcolsep{3pt}
\begin{center}
\begin{tabular}{l| r r r r|}
\hline
{\bf Variable}	&	Comp.1     &    Comp.2   &   Comp.3 & Comp.4(ref) \\ \hline
Location($\widehat{\zeta}_{k})$  & -3.696 & -0.978 & 1.108 & 1.468\\ \hline
Intercept &	-0.162 &-10.706 &-2.294 & \\
Year	  &	 0.006 & -0.076 & 0.037 & \\
Age	      &	 0.001 &  0.096 &-0.0289 & \\
Grade	  &	 0.119 (**) &  0.909 (**) & 0.103 (**) & \\
Not smsa  &	 0.689 (**) &  1.461 (**) & 0.168 & \\ 
South	  &	-1.044 & -6.566 & 0.127 & \\
South*Year&	 0.017 & -0.036 & 0.003 & \\ \hline
\end{tabular}
\end{center}
\caption{Union data. Prior Probability Model, parameter estimates. BIC-based solution, $K=4$ components. $(*)$ and $(**)$ denote \emph{weakly} ($0.1\leq p \leq 0.05$) and \emph{strongly} ($p \leq 0.05$) significant effects, respectively.} \label{RIS_pi_BIC}
\end{table}

We now proceed to commenting the estimates of parameters in the models for the prior probabilities of component membership, in order to have some comparison with QP approaches.  The two solutions, obtained by using the AIC and the BIC for choosing the number of components, respectively, agree on most strongly significant effects, $Grade$ (in three out of three and in four out of five components in BIC and AIC based solutions), $NS$ (in two out of three and in one out of five components in BIC and AIC based solutions). In the AIC based solution we have also found \emph{strongly} significant effects for $Year$ and $Age$.
The impact of individual covariates means may be better understood if we look at the component-specific intercept estimates $\zeta_k$, $k=1,\dots,K$; for example, by looking at the estimates for $Grade$, we may observe that these are significant and high in (absolute) magnitude in component 5 of the AIC-based solution (associated to an estimate $\widehat{\zeta}_{2}= 4.128$) and in component 3 of the BIC based solution (associated to an estimate $\widehat{\zeta}_{2}= 1.108$). Both intercept estimates represent the maximum value of the estimated random effect distribution, and this suggests a positive \emph{between} effect estimate for $Grade$ as observed in the $QPPar$ specification. The same argument applies to $NS$ whose effect is strongly significant for component 2 membership (with a substantially high but negative intercept estimate) in the AIC-based solution, while the interpretation of the same variable when the BIC is employed is not that clear-cut. This may suggest, for the BIC-based solution, some issues with the logit model estimates for parameters in the model for prior probabilities and component 2, as the very high (but negative) estimate for the corresponding intercept may suggest some numerical instability. So, one may wonder whether the random effect distribution estimate that we obtain by choosing the number of components by BIC may bee too inaccurate, when compared to that obtained by using the AIC. 

\section{Concluding remarks}\label{conclusions}
In this paper, we have described a semiparametric approach to deal with covariates endogeneity in random effect models for longitudinal responses. While we focus on the simplest case of discrete time and common measurement occasions, the proposed approach can be readily extended to other scenarios as well, that is studies in continuous time with (at least partially) non common and unequally spaced time occasions. Moreover, while we discuss, for sake of simplicity, only balanced designs, the approach works just as fine with unbalanced studies.

As it has been shown by using standard arguments, in the presence of potential endogeneity, the bias can be handled by partitioning, conditioning (both in a FE and RE fashion) or joint modelling in a random effect framework. The random effect model \emph{per se} does not necessarily lead to non consistent estimators; rather, in this respect, it is the adopted parametrization that plays a fundamental role. Along the lines of the \emph{auxiliary} regression by \cite{mund:1978}, a general class of models to deal with several well known issues can be defined; just to give a few examples, we may consider the initial conditions problem, see e.g. \cite{aitk:alfo:1998}, \cite{wool:2005} and \cite{rabe:skro:2014} for a thourough review, the handling of models with pre-treatment baseline values, see \cite{alfo:aitk:2006}, the pattern mixture models for monotone missing data proposed by \cite{roy:2003} and, in general, any random effect models in the presence of \emph{endogenous} covariates. We have started the discussion from the linear regression model and extended it to the generalized linear regression models. However, we may be interested to make a few steps ahead to define finite mixture approaches to robust regression modelling, such as quantile and Mquantile regression models, as in \cite{alfo:etal:2016}.
The simulation study shows that the proposed approach offers in all analysed conditions good performance in terms of both bias and MSE of model parameter estimates; further, when we move from the linear to the nonlinear (Bernoulli case with probit link) it outperforms all the others, therein included the fixed effect and the bias corrected fixed effect estimator.

\section*{Conflicts of interest}
The Authors declare that they have no conflict of interest

\bibliographystyle{plainnat}
\bibliography{biblio}
\end{document}